\documentclass{aa}  
\usepackage[varg]{txfonts}
\usepackage{graphicx}
\graphicspath{{Figures/}}
\usepackage{txfonts}
\usepackage{float}
\usepackage{hyperref}
\usepackage{multirow}
\usepackage{color}
\usepackage{csquotes}
\usepackage{caption}
\newcommand{\cla}[1]{#1}

\begin{document}

%Define shorthands for a few commonly refereed papers 
\defcitealias{feiden:2013}{FC13}
\defcitealias{macdonald:2014}{MM14}
\defcitealias{delfosse:1999}{D99}
\defcitealias{ribas:2003}{R03}
\defcitealias{wilson:2017}{W17}

\title{Multi-scale magnetic field investigation of the M-dwarf \\ eclipsing binary CU Cancri 
}

\subtitle{}

   \author{A. Hahlin\inst{1} \and O. Kochukhov\inst{1} \and A.D. Rains\inst{1}\and J. Morin\inst{2} \and G. Hussain\inst{3} \and L. Hebb\inst{4,5} \and K. Stassun\inst{6}} 
   \institute{Department of Physics and Astronomy, Uppsala University, Box 516, SE-751 20 Uppsala, Sweden \\\email{axel.hahlin@physics.uu.se}
   \and
   LUPM, Universit\'e de Montpellier, CNRS, Place Eug\`ene Bataillon, F-34095 Montpellier, France
   \and
   European Space Agency, Keplerplaan 1, 2201 AZ Noordwijk, Netherlands
   \and
   Physics Department, Hobart and William Smith Colleges, 300 Pulteney Street, Geneva, NY 14456, USA
   \and
   Department of Astronomy, Cornell University, 245 East Avenue Ithaca, NY 14850, USA
   \and
   Department of Physics and Astronomy, Vanderbilt University, Nashville, TN 37235, USA
   }

    \date{Received: 2023-11-28/ Accepted: 2024-01-31}
    
\abstract
{}%Context(optional)
{We aim to characterise the magnetic field of the eclipsing binary CU Cancri 
%(CU Cnc) 
consisting of two M-dwarf components. The determination of magnetic field parameters of this target enables comparisons with both observations of similar stars and theoretical predictions of the magnetic field strength specifically in CU Cnc. The target is therefore providing an excellent opportunity to test our understanding of the generation of magnetic fields in low-mass stars and its impact on stellar structure.}%Aims
{We use spectropolarimetric observations 
obtained with ESPaDOnS at the CFHT
to investigate the magnetic properties of CU Cnc. In order to improve the signal, we use least-squares deconvolution (LSD) to create average line profiles. From these LSD profiles, we extract information about the radial velocities of the components, significantly expanding the number of radial velocity measurements available and allowing for a determination of the orbital parameters. Stokes $V$ LSD profiles are used with Zeeman Doppler imaging to obtain large-scale magnetic field structures on the two components. We also use detailed polarised radiative transfer modelling to investigate the small-scale fields by utilising Zeeman splitting of magnetically sensitive \ion{Ti}{I} lines in non-polarised spectra.}%Methods 
{We obtain both small- and large-scale magnetic field properties for the two components. The large-scale fields are dominantly poloidal and have an average strength of approximately 100\,G on both components. This analysis of the large-scale fields likely suffers from some amount of hemisphere degeneracy due to the high inclination of the target, which would cause the large-scale field strength of the components to be underestimated. Both components also show unusual magnetic field configurations compared to stars with similar parameters, the primary is weakly axisymmetric ($\sim$\,10\,\%) and the secondary has a strong torroidal contribution ($\sim$\,20\,\%). The small-scale fields are significantly stronger, at 3.1 and 3.6\,kG for the primary and secondary respectively. This measurement is in excellent agreement with surface field strength predictions for CU Cnc from magnetoconvective stellar evolution models. These results indicates that magnetic fields could play a significant role in the radius inflation due to convective inhibition. }%Results
{} %Conclusions(optional)


\authorrunning{A. Hahlin et al.}
\titlerunning{Multi-scale magnetic field investigation of the eclipsing binary CU Cnc}
\maketitle
%
%-------------------------------------------------------------------
\section{Introduction}
\label{sec:introduction}

Magnetic fields are ubiquitous on partially- and fully-convective stars as the convective motions are thought to be the primary generator of stellar magnetic fields in cool stars \citep[e.g.][for a review]{charbonneau:2014}. It is possible to study these fields using several different methods. One of the most commonly utilised method known as Zeeman Doppler imaging \citep[ZDI, see][for a review]{kochukhov:2016},
uses a time series of spectropolarimetric observations in order to obtain the magnetic field structure on the stellar surface. This method does, however, suffer from significant signal cancellation as polarisation signals of opposite polarities in nearby surface elements might cancel each other out. A consequence of this is that most of the smaller magnetic structure of the surface is not observable with this technique. In order to mitigate this shortcoming and quantify the small-scale fields, Zeeman broadening or intensification \citep[e.g.][]{reiners:2012} is used on non-polarised spectra. These diagnostic methods are only sensitive to the absolute strength of the magnetic field, avoiding the field cancellation present in ZDI studies relying on polarised observations. Although somewhat dependent on stellar mass \citep[e.g.][]{morin:2010,vidotto:2014}, studies have found that the magnetic fields observed with methods sensitive to the small-scale fields consistently give field strengths about one order of magnitude stronger than the large-scale fields for most stars \citep[e.g.][]{lavail:2019,see:2019,kochukhov:2021}. 

The different spatial scales also govern different dynamics and interactions in and around late-type stars. The large-scale field will reach out beyond the photosphere and interact with the stellar surroundings, including planets \citep[e.g.][]{carolan:2021}. As M-dwarfs are popular targets of exoplanet searches, characterising their magnetic field is important to understand both the host stars themselves as well as the space weather environment of any short-period planets. In the case of binaries, the magnetic field could connect the two components, leading to magnetic interaction changing along the orbital phase \citep[e.g.][]{gregory:2014,Pouilly:2023}. Since the small-scale fields are stronger, they are more prevalent in the formation of local structures on the stellar surface, such as starspots that can host very strong fields \citep[e.g.][]{okamoto:2018}. Another aspect where the magnetic field might play a major role is the inhibition of convection within a convective zone of a star \citep{mullan:2001}. This causes an inflation of the stellar radii, resulting in systematic mismatches between observations and theoretical predictions \cla{typically around a few percent for M-dwarfs but ranging up to and above 10\% in some cases \citep[e.g.][]{lopez-morales:2007,parsons:2018,morrell:2019}. The possible connection to magnetic fields was illustrated by \cite{lopez-morales:2007} who found a correlation between activity indicators and the radius discrepancy.}

Binary stars are key objects in the investigation of this issue. Eclipsing binaries in particular allow an accurate determination of the stellar parameters, such as masses obtained by studying radial velocity shifts due to gravitational interaction between the two components and radii obtained during the eclipse when the brightness of the binary temporarily decreases. Analysis of these two interactions enables model-independent determination of these key stellar parameters with an uncertainty of a few \% \citep[see][for a review of binary stars with accurate parameters]{torres:2010} which in turn puts strong constraints on stellar models. This has revealed a radius discrepancy where many stars show larger radius than predicted, possibly resulting in an incorrect determination of stellar ages \citep[e.g.][]{popper:1997}.

One of the spectroscopic binaries included in \cite{torres:2010} is CU Cancri (GJ 2069A,  	2MASS J08313759+1923395), a mid M-dwarf binary containing two similar-mass components with the stellar masses slightly above the fully convective limit of $0.35\,M_\odot$ \citep{chabrier:1997}. The star has been investigated in the past \citep[e.g.][herefater D99, R03 and W17]{delfosse:1999,ribas:2003,wilson:2017}, and as such its fundamental stellar parameters are well known (see Table~\ref{tab:Stellar_Param} for a selection relevant to this work). The binary is known to be active as \citetalias{ribas:2003} reported an X-ray luminosity from the ROSAT survey \citep{voges:1999} close to the dynamo saturation limit. It is also one of the stars where the observed radius appears to not match theoretical prediction, \citetalias{delfosse:1999} reported that CU Cnc has a spectral class that is too late for its mass. While this discrepancy could be due to many effects, the components of CU Cnc follow the trend presented by \cite{lopez-morales:2007}, \cla{showing increased radii by about $8\pm3$\,\% compared to models. The fact that the system follows the trend suggests that there is a connection between the radius inflation and stellar magnetic activity.}
%the components of CU Cnc follow the trend presented by \cite{lopez-morales:2007}, which indicates a connection between its radius and activity.

To this end, magnetic fields have been suggested as a possible cause for the radius anomaly and the effect of magnetic fields have been included in the stellar evolution models to test this idea \citep[e.g.][hereafter FC13 and MM14]{feiden:2013,macdonald:2014}. These works found significantly different surface magnetic field strengths, with \citetalias{feiden:2013} predicting field strengths of a few kG while \citetalias{macdonald:2014} estimating field strengths of a few hundred G. When comparing these results with the observational studies of other binary systems by \cite{kochukhov:2019} and \cite{hahlin:2021}, there is a good agreement between the measured small-scale field strengths and the predicted surface fields from \citetalias{feiden:2013}. Still, until now only a small number of binary stars have had their magnetic field characterised both theoretically and observationally.

The only binary star investigated by \citetalias{feiden:2013} that has yet to receive an observationally characterised magnetic field is CU Cnc. A detailed characterisation of the magnetic fields on the components of CU Cnc would therefore expand the sample of eclipsing binary stars with both observationally constrained and theoretically predicted magnetic field parameters. In addition, CU Cnc also has a predicted magnetic field from \citetalias{macdonald:2014}. As the field strengths of \citetalias{feiden:2013} and \citetalias{macdonald:2014} are significantly different, observational analysis would therefore further showcase which of these models best predict the surface magnetic field strengths on binary stars.

With a magnetic field characterisation, it will also be possible to compare the magnetic field of CU Cnc with other studies of the large- \citep[e.g.][]{morin:2008} and small-scale \citep[e.g.][]{shulyak:2019} magnetic field on the surfaces of other M-dwarfs. This can give an indication of any systematic differences present in the generation of magnetic fields on binaries or trace the behaviour close to the fully convective limit.

Another interesting aspect to investigate in the context of magnetic field on binary stars is to use the fact that the two components of CU Cnc have rather similar masses. As binaries form simultaneously and in the same region, there should be close similarity between the two components. As magnetic field generation is believed to be caused by convection (dependent on stellar parameters) and rotation (synchronised in close binary systems), there should be little difference in the magnetic properties of the two stars. Investigating the similarities and differences between the two components is a good way to quantify how predictable magnetic field parameters could be in other contexts where the similarity of stars are less apparent. Other works looking at binary stars with similar mass components have found that the large-scale fields obtained with ZDI are often significantly different between the two components \citep[e.g.][]{donati:2011,kochukhov:2017,rosen:2018,kochukhov:2019,lavail:2020,Pouilly:2023} while the small-scale fields from Zeeman broadening or intensification are found to be more similar \citep[e.g.][]{kochukhov:2019,hahlin:2022,Pouilly:2023}. This difference between spatial scales is likely a consequence from the inherent evolution of the large-scale fields found in single stars \citep[e.g.][]{boro-saikia:2018}, but also indicates a weaker variability in the small-scale fields.

Section~\ref{sec:obs} covers the data used in this work as well as the treatment to extract individual spectra of the two components of CU Cnc. Application of the least-squares deconvolution method is described in Sect.~\ref{sec:LSD} as well as the determination of the orbital solution and magnetic signatures. In Sect.~\ref{sec:metallicity}, we discuss the chemical composition of CU Cnc, both the overall metallicity as well as a claimed lithium detection by \citetalias{ribas:2003}. Section~\ref{sec:LS} and \ref{sec:SS} are dedicated to the determination of the large- and small-scale surface magnetic fields on the two components. The results from Sect.~\ref{sec:LS} and \ref{sec:SS} are discussed in Sect.~\ref{sec:discussion} and concluding remarks are made in Sect.~\ref{sec:conclusions}.

\section{Observations}
\label{sec:obs}

\subsection{ESPaDOnS spectra}
A series of 20 high resolution spectra, obtained with ESPaDOnS \citep{donati:2003} at the CFHT \footnote{Data obtained from \url{https://www.cadc-ccda.hia-iha.nrc-cnrc.gc.ca/en/}}, is used for this work. ESPaDOnS is an optical spectropolarimeter with a wavelength coverage between 3600 and 10000\AA\ and a resolving power of $\approx$65000. The observations were taken over the two weeks between January 3rd and 16th, 2012. Each observation consisted of four 1140\,s sub-exposures obtained with a different polarimeter configuration in order to derive circularly-polarised while minimising instrumental artefacts. Here we use the Stokes $I$ spectra from individual 1140\,s sub-exposures to study radial velocity variation at high time resolution. We use the combination of four sub-exposures (i.e. 4560\,s effective exposure time) for the analysis of polarisation profiles. The same sequence of four sub-exposures is also used to derive the null polarisation spectra in which stellar signal is cancelled out.

The data reduction was performed using the \texttt{Libre-ESpRIT} package \citep{donati:1997}. During most nights, the reduced Stokes $I$ and $V$ observations have a signal-to-noise ratio (S/N) of approximately 130 and 260 per pixel respectively in the wavelength region between 9600 and 9800 \AA\, that contains the \ion{Ti}{I} lines used in Sect. \ref{sec:SS}. Information on individual observations, including the heliocentric Julian dates of mid-exposures and S/N values, can be found in Table~\ref{tab:rv} and \ref{tab:fap} for the intensity and polarisation data respectively.

\subsection{Spectral disentangling}
\label{sec:specDis}

\begin{figure}
    \centering
    \includegraphics[width=\linewidth]{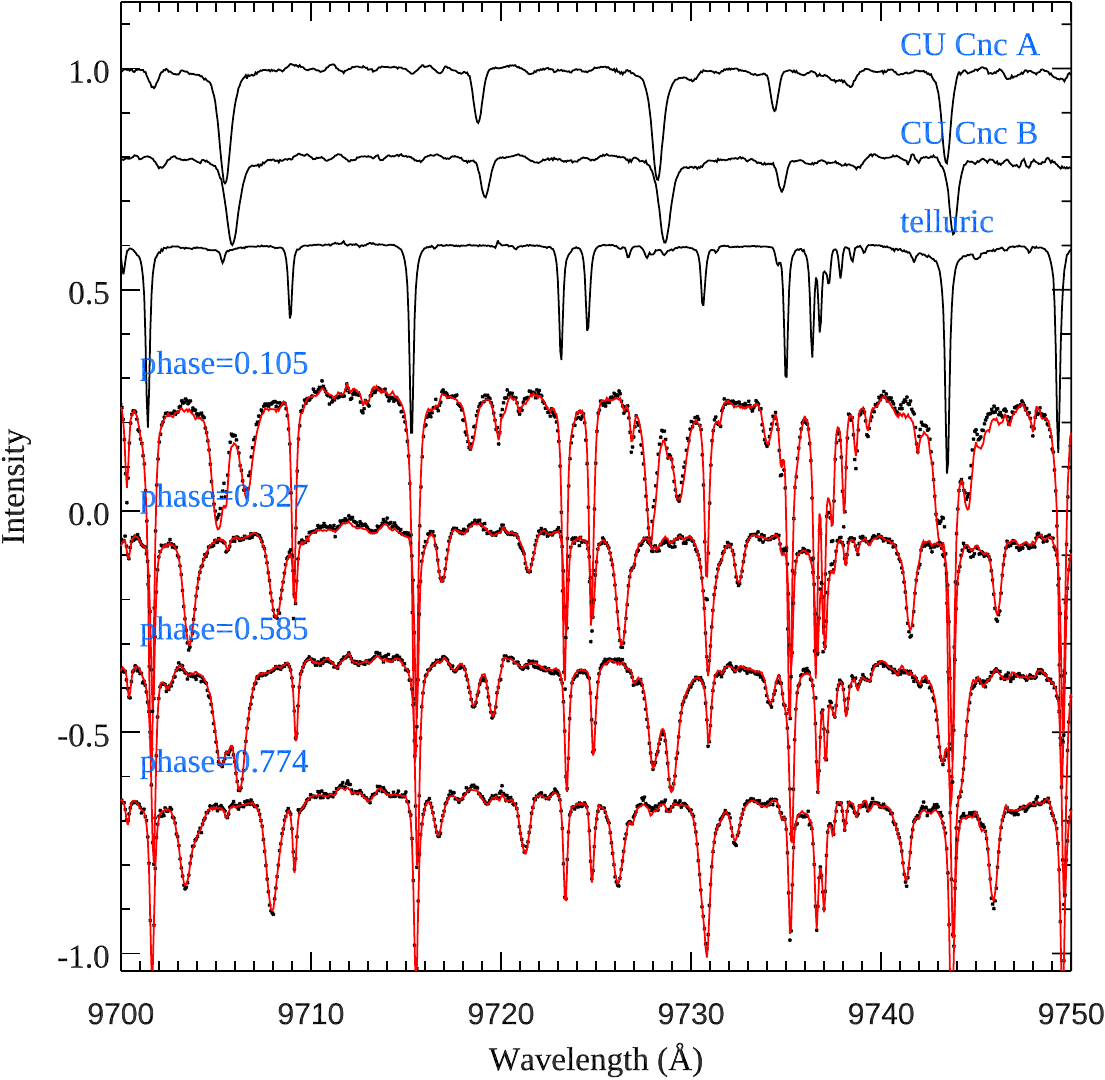}
    \caption{Disentangled spectra of CU Cnc. The top three spectra show the time averaged spectra of the A and B component as well as the telluric absorption. The bottom four compare the observed spectra (black dots) and the model fit with a combination of the three top spectra (red line).}
    \label{fig:spec_dis_ti}
\end{figure}

In order to study the properties of the binary components of CU Cnc individually, we need to separate spectra of the two components. This is achieved by using the spectral disentangling method described in \cite{folsom:2010} with the additional functionality of simultaneously disentangling the telluric signal \citep[see][]{kochukhov:2019}. The method assumes that a time series observation of a spectroscopic binary can be described by three spectral components, one for each binary component and one for the telluric absorption. The stellar contributions are assumed to be shifted in velocity but constant in time. The telluric component is scaled to match individual observations. The stellar components are given radial velocities based on the orbital solution obtained in Sect.~\ref{sec:orbit}. In this study we disentangled a region between 9630 and 9820 \AA\, containing the \ion{Ti}{I} lines of interest for the magnetic field investigation covered in Sect.~\ref{sec:SS}. The obtained time-averaged spectra of the two components and the telluric absorption spectrum can be seen in Fig.~\ref{fig:spec_dis_ti} for a selection of phases. We also attempted to disentangle spectra further to the blue in order to perform a metallicity analysis (see Sect.\,\ref{sec:metallicity}), but found that the method has difficulty in correctly recovering the depth of wide molecular bands as the overlap of many molecular lines makes it impossible to distinguish between contributions from the two components. 

\section{Least-squares deconvolution}
\label{sec:LSD}

In order to reliably detect circular polarisation signals in stars the use of multi-line techniques are needed. To this end, we used least-squares deconvolution \citep[LSD,][]{donati:1997,kochukhov:2010}, which combines information from a large number of lines in order to increase the S/N. We use the VALD database \citep{ryabchikova:2015} to generate an atomic line mask between 4500 and 9850\,\AA\,assuming stellar parameters of T$_\mathrm{eff}=3\cla{5}00$\,K and $\log g=5.0$. We also remove wavelength regions containing telluric absorption or strong stellar lines. The minimum central line depth was set to 0.2 yielding a total of 2681 lines in the mask. For the mask we select average line parameters corresponding to a wavelength of $\lambda_0=6450$\,\AA,  a mean Land\'e factor of $\bar{g}=1.2$, and a depth of 0.5. These parameters were chosen to closely correspond to the average parameters of lines present in our LSD mask of CU Cnc.

Using this mask, we first generate Stokes $I$ profiles from individual polarisation sub-exposures in order to get radial velocity measurements for the orbital solution (see Sect. \ref{sec:orbit}). We then generate Stokes $I$ and $V$ from the polarised observations. For the Stokes $V$ profiles we use a mask that does not exclude wavelengths with tellurics as these absorption features are not polarised. This allows us to use more lines in the mask in order to increase the S/N. For this mask the total number of lines are 2952. Inspecting Stokes $V$ LSD profiles generated from the masks with and without tellurics shows no major difference in the shape or strength, indicating that the magnetic information contained within the spectra should not be significantly modified by choice of mask in this case. Typical improvements in S/N of the LSD profiles are about a factor of 40 compared to observations in individual spectral lines.

\subsection{Orbital solution}
\label{sec:orbit}

The orbital solution is a prerequisite to model the binary nature of CU Cnc. In order to obtain the orbital solution we first measure individual radial velocities from the Stokes $I$ LSD profiles by fitting two Gaussians to each profile. The resulting radial velocities are reported in Table \ref{tab:rv}. We then used the obtained radial velocities from our spectra, as well as those reported by \citetalias{delfosse:1999} and \citetalias{wilson:2017}, to fit the orbital parameters of CU Cnc using an initial guess based on \cla{the results from \citetalias{delfosse:1999}.}

Before obtaining the final orbital parameters, we initially investigated the possibility of a non-circular orbit, but found the eccentricity to be compatible with a circular orbit. For this reason, the eccentricity parameter was omitted from the final orbital solution. This is in line with previous investigations of CU Cnc's orbit, as no eccentricity has been reported by either \citetalias{delfosse:1999} or \citetalias{wilson:2017}. Another aspect to explore is that \citetalias{wilson:2017} also reported a small period variability. This was investigated by comparing the period variability using the description from \cite{wilson:2005} obtained from different subsets of the radial velocity data. This was done by using either all or our data and the data from \citetalias{delfosse:1999} and \citetalias{wilson:2017} or a subset of two different observation sequences. We find that fitting radial velocity data with the period change prescription from \cite{wilson:2005} yields somewhat different results depending on which datasets are used.
%, resulting in slightly different periods. 
The finding of a period variability in \citetalias{wilson:2017} could be due to the fact that the radial velocities from \citetalias{wilson:2017} suffer from higher uncertainties compared to those from \citetalias{delfosse:1999} and this work. We find that a better fit can be obtained adopting different centre-of-mass radial velocity $V_{\gamma}$, which we do instead of assuming a period variability. This is probably justified by the fact that CU Cnc has a visual companion \citep{giclas:1959,delfosse:1999} that could cause a small variation of $V_\gamma$ %Comment on the secondary component

\begin{figure*}
    \centering
    \includegraphics{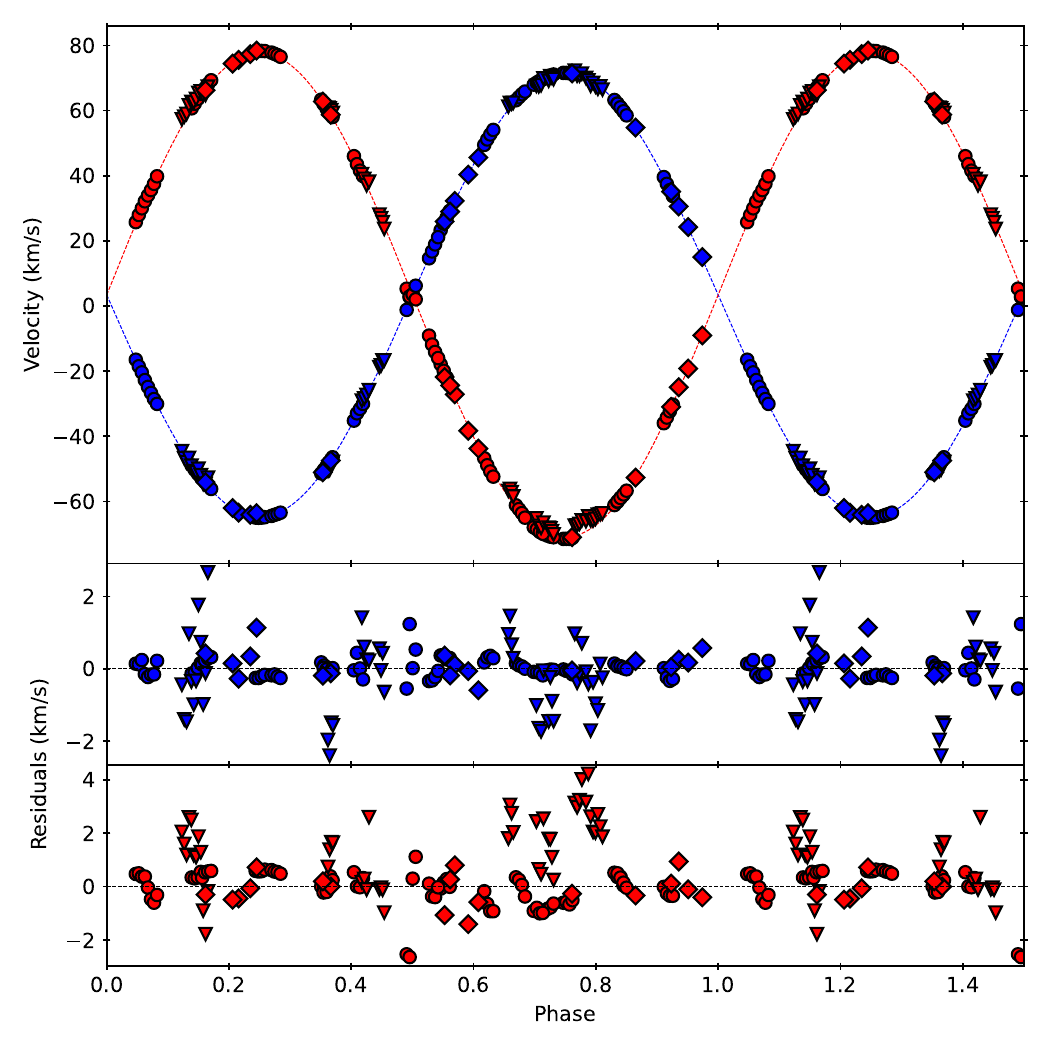}
    \caption{\textit{Top:} orbital solution for CU Cnc, shown as a function of the orbital phase. Circles mark the individual radial velocities obtained from the LSD profiles in this work (see Table\,\ref{tab:rv}) while triangles and diamonds correspond to the measurements taken from \citetalias{wilson:2017} and \citetalias{delfosse:1999} respectively. Dashed lines show the best-fit orbital solution. The blue and red colours refer to the primary and secondary component respectively. \textit{Bottom:} Residuals between the individual radial velocity measurements and the orbital solution for the primary (upper panel) and secondary (lower panel) component.}
    \label{fig:orbit}
\end{figure*}

The resulting fit can be seen in Fig.~\ref{fig:orbit}. The obtained orbital parameters are given in Table~\ref{tab:Stellar_Param}. Comparing with the orbital solutions presented by \citetalias{delfosse:1999} and \citetalias{wilson:2017}, we find good agreement. We do note systematically larger residuals of the \citetalias{wilson:2017} data-points in the lower half of Fig~\ref{fig:orbit}. While the $V_\gamma$ for our observing epoch is slightly different from the ones reported in previous works, our best fit value for the \citetalias{delfosse:1999} epoch is $4.38\pm0.07$ km\,s$^{-1}$ which does agree with the earlier estimates of $V_\gamma$. 

\begin{table}
    \centering
    \caption{Stellar and orbital parameters for CU Cnc.}
    \label{tab:Stellar_Param}
    \begin{tabular}{lrr}
        \hline\hline
         & CU Cnc A & CU Cnc B \\ %Both are spectral type M3.5 Ve
        \hline
        Mass ($M_\odot$) & 0.4358(8) & 0.3998(14) \\ 
        Radius ($R_\odot$)$^*$ & 0.4317(52) & 0.3908(94) \\
        $\log g^{\cla{*}}$ & 4.804(11) & 4.854(21) \\
        $T_{\mathrm{eff}}$ (K)$^{*}$ & 3160(150) & 3125(150) \\ 
        $t_0$ (HJD) & \multicolumn{2}{r}{2455477.06800(35)} \\
        $P$ (d) & \multicolumn{2}{r}{2.77146871(34)} \\
        $V_{\gamma\cla{,\mathrm{ESPaDOnS}}}$ (km\,s$^{-1}$)$^{**}$ & \multicolumn{2}{r}{3.411(38)}\\
        $V_{\gamma,\text{\citetalias{wilson:2017}}}$ (km\,s$^{-1}$)$^{**}$ & \multicolumn{2}{r}{\cla{3.113(99)}}\\
        $V_{\gamma,\text{\citetalias{delfosse:1999}}}$ (km\,s$^{-1}$)$^{**}$ & \multicolumn{2}{r}{\cla{4.376(67)}}\\
        $V$ (km\,s$^{-1}$)& 68.195(44) & 74.33(12) \\
        $a$ ($R_\odot$) & \multicolumn{2}{r}{7.800(7)} \\
        $i$ (deg)$^*$ & \multicolumn{2}{r}{86.34(3)}\\
        \hline
    \end{tabular}
    \tablefoot{$^*$corresponds to the parameters obtained by \citetalias{ribas:2003}, other parameters are obtained using the orbital solution in this study. \cla{$^{**}$\,$V_{\gamma}$ refers to the centre-of-mass velocity during each epoch of observations.}}
\end{table}

\subsection{Magnetic signatures}
\label{sec:FAP}

\begin{figure}
    \centering
    \includegraphics[height=0.75\textheight]{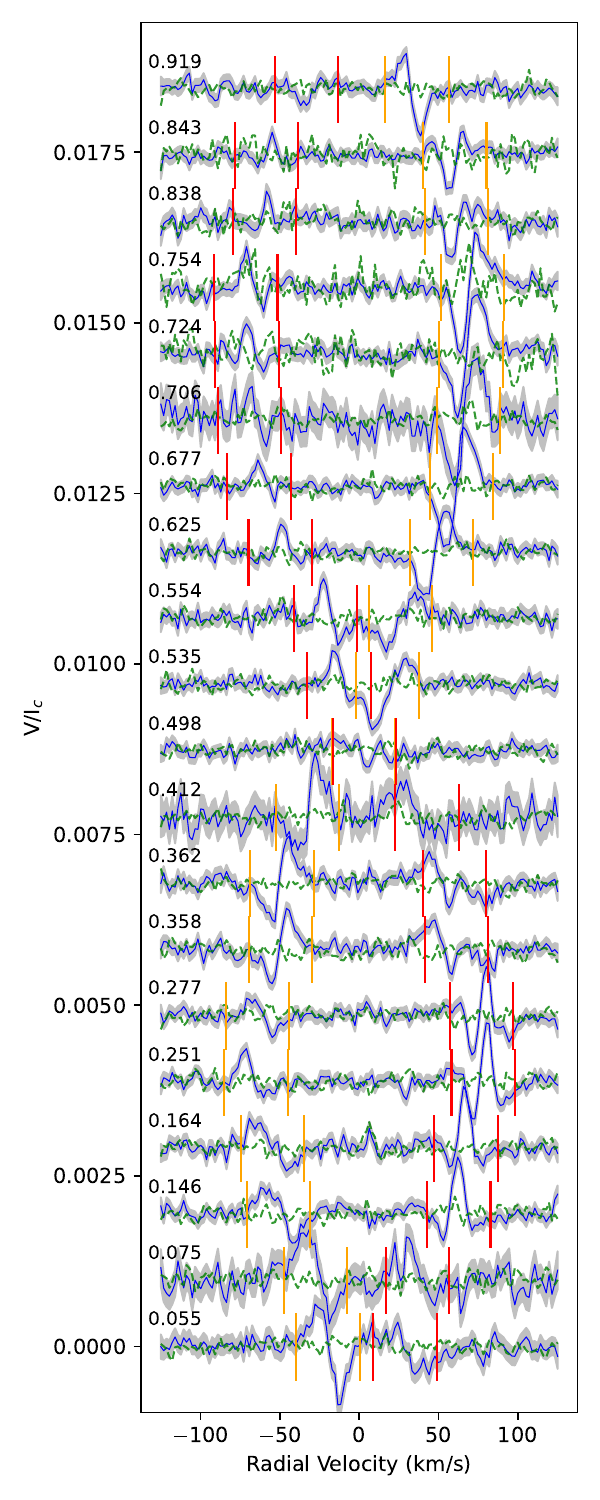}
    \caption{Observed Stokes $V$ LSD profiles. The shaded grey area represents the uncertainty of each LSD profile. The null polarisation profiles are also shown with the dashed green lines. The vertical bars indicate velocity intervals for calculation of the FAP and $\langle B_{z}\rangle$ for the primary (orange bars) and secondary (red bars). The spectra are offset vertically according to the orbital phase. The phases for each observation are shown on the left-hand side and are calculated from the orbital solution derived in Sect.~\ref{sec:orbit}.}
    \label{fig:stokesV}
\end{figure}
A common way to evaluate the presence of any magnetic signal is to utilise the false alarm probability \citep[FAP,][]{donati:1997} and calculate the longitudinal magnetic field $\langle B_z\rangle$. Both of these characteristics can be obtained from the Stokes $V$ LSD profiles. As each LSD profile contains signals from the two components, we use the obtained radial velocities from Sect.~\ref{sec:orbit} of each component to evaluate the FAP and $\langle B_z\rangle$ in a velocity window of $\pm20$\,km\,s$^{-1}$ from the measured radial velocity.

The FAP is determined by calculating the probability that any signal in the Stokes $V$ and null profiles could be described with a straight line at zero (null hypothesis). Depending on the value of the FAP, the signal is either labelled a definite detection (DD), marginal detection (MD), or non-detection (ND). These thresholds are FAP $<10^{-5}$ for DD, FAP $<10^{-3}$ for MD, and FAP $>10^{-3}$ for ND as defined by \cite{donati:1997}. The resulting FAPs can be seen in Table~\ref{tab:fap}. The primary has a detected signature at almost every observation, with only the phase 0.277 yielding a marginal detection and phase 0.498 giving a non-detection (likely due to signal overlap with the secondary). Overall, the secondary exhibits weaker magnetic field signals, with 12 detections (9 of which are definitive) and 8 non-detections. All but one null profile show no significant magnetic field signals. The exception is the phase 0.754 where there is a detection in the null profile for both the A and B component.

The longitudinal magnetic field is calculated from the first-order moment of the Stokes $V$ profile \citep[see e.g.][]{donati:1997,kochukhov:2010}. A complete list of $\langle B_z\rangle$ measurements is provided in Table~\ref{tab:fap}. We find longitudinal magnetic field strengths on the order of 100\,G for both components.
The primary exhibits longitudinal fields above 100\,G around the 0 phase and between the phases 0.4 and 0.75.
The longitudinal field of the secondary appears more concentrated at specific epochs, with 100\,G fields only appearing around phase 0 and 0.4.
The uncertainties of our $\langle B_z\rangle$ measurements also confirm the more robust detection of magnetic signal on the primary component as they tend to be lower than the secondary.

\cla{\section{H$\alpha$ emission}
\label{sec:emission}
The spectra of CU Cnc show a strong emission in H$\alpha$. 
This has been reported before (e.g. \citetalias{ribas:2003} and \citetalias{wilson:2017}), but the new time-resolved observations available in our study provide a unique opportunity to investigate potential time dependence and systematic difference in the H$\alpha$ emission of the two components. As reported in previous work, the emission from the two components is well separated in wavelength, except near the eclipses, making it possible to measure individual emissions for the two stars. We calculated the H$\alpha$ equivalent width of the two components at all phases when the radial velocity difference was sufficient to clearly separate the two emission features. We note that even with this criterion we are not able to use the full width of the H$\alpha$ line \citep[e.g.][]{schoefer:2019} as the wings of each emission feature are still blended. The primary component shows a slightly stronger emission compared to the secondary during all but one observation at phase 0.146. We obtained median H$\alpha$ emission equivalent widths of $2.5$ and $2.1$\,\AA\, for CU Cnc A and B, respectively. There is also a scatter of about 0.3 and 0.5\,\AA\, for the same components meaning that even if the primary's H$\alpha$ appears stronger at almost every phase, the difference with the secondary is not particularly significant. Similar to what was found by \cite{tsvetkova:2023} for the M-dwarf binary FK~Aqr, H$\alpha$ emission changes in CU~Cnc system is not coherent with the rotational phase
suggesting that emission regions are not associated with specific stable surface features.

The components also show a double-peaked emission profile. This has been previously reported for CU Cnc (\citetalias{wilson:2017}) and other active M-dwarfs \citep[e.g.][]{stauffer:1997,tsvetkova:2023}. We find similar separations of about 0.7\,\AA\, compared to \citetalias{wilson:2017}. The double-peaked emission has been shown to be caused by an optically thick chromosphere \citep{worden:1981,stauffer:1986} with non-thermal velocity fields \citep{cram:1985}. There are, however, two observations at phases 0.146 and 0.358 when CU Cnc B does not show a double-peaked emission. This indicates a temporary deviation from these conditions. 
}

\section{Metallicity}
\label{sec:metallicity}

\subsection{[Fe/H]}
Determining the metallicity of M dwarfs from optical spectra is challenging. The density of both atomic and molecular lines makes continuum placement very difficult while simultaneously requiring accurate line parameters for a large number of lines. Regardless, some investigations into the metallicity of CU Cnc have been done before. From optical photometry, \citetalias{delfosse:1999} argued for a supersolar metallicity based on the late spectral class compared to the masses of the components. This was challenged by \citetalias{ribas:2003}, who used the spatial velocities to associate CU Cnc with the Castor moving group which would indicate solar metallicity. However, the usefulness of the Castor moving group as a metallicity indicator has been put into question \citep[e.g.][]{mamajek:2013} as its members have a rather large velocity scatter and are therefore unlikely to originate from the same point in the Galaxy. Using spectroscopy, \citetalias{wilson:2017} analysed a single spectral line, \ion{Fe}{I} 8611.8\,\AA, in a region with relatively low line density and found a metallicity of [Fe/H]$=+0.4$.

While it seems that CU Cnc is likely a metal-rich star, there are still some complications to consider. Magnetic fields could impact the results of both \citetalias{delfosse:1999} and \citetalias{wilson:2017}. 
\cla{In the first case, convective inhibition can increase the stellar radii of the two components introducing systematic errors in the photometric analysis. In the second case, the \ion{Fe}{I} 8611.8 line has a Land\'e factor of 1.49 and could therefore be affected by strong magnetic fields due to the increase in equivalent width caused by Zeeman splitting \cite[e.g.][]{basri:1992}.}
Investigating the sensitivity of this feature to the Zeeman intensification by calculating synthetic spectra with different field strengths, we find that the \ion{Fe}{I} 8611.8\,\AA\,line is moderately sensitive to magnetic fields. At a field strength of 1\,kG the change in the equivalent width is about 3 \% and at 3\,kG it reaches about 10 \%. For the 3\,kG case, the equivalent width increase of 10\,\% due to magnetic fields would correspond to a reduction in abundance of about 0.1. As the components exhibit magnetic signatures in Stokes $V$ (Sect.\,\ref{sec:FAP}) and a strong H$\alpha$ emission, reported by both \citetalias{ribas:2003} and \citetalias{wilson:2017}, it is not unlikely that the magnetic field could introduce biases in the metallicity determination if left unaccounted for.

For this reason, we apply a different approach to finding metallicity of the system by comparing the observed spectra of CU Cnc with observations of other M dwarfs with metallicities either determined from hotter stellar companions or calibrated from such binary systems. We make use of a grid of low-resolution optical spectra of cool dwarf benchmark stars observed by \cite{zerjal:2021} and \cite{rains:2021}. These stars were observed with using the WiFeS instrument \citep[Wide Field Spectrograph,][]{dopita:2007} on the ANU 2.3\,m Telescope at Siding Spring Observatory, Australia, and were reduced using the PyWiFeS pipeline \citep{childress:2014}. We make use of the flux calibrated R$\sim$7000 red arm spectra ($5400 - 7000$\AA) for our comparison, which have median S/N$\sim$130 and a substantial wavelength overlap with our ESPaDOnS data, and use the literature stellar parameters compiled in \cite{rains:2021}.

\begin{figure}
    \centering
    \includegraphics[width=\linewidth]{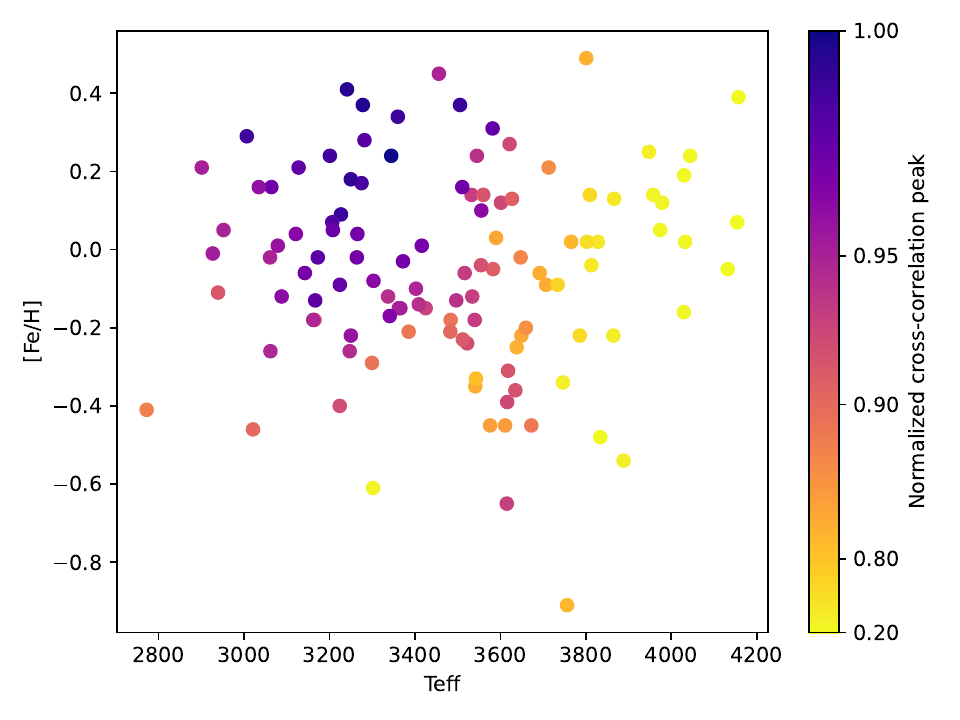}
    \caption{Strength of the normalised cross-correlation in [Fe/H]-$T_{\mathrm{eff}}$ space. Each point represents the cross-correlation peak of the spectra of one benchmark star.}
    \label{fig:ccf}
\end{figure}

In order to make the different data sets consistent we reduce the resolution of our ESPaDOnS spectra to that of the WiFeS's spectral resolution and then interpolate them to the same wavelength grid. We then use a Gaussian smoothing method described in \cite{ho:2017} on both the ESPaDOnS and WiFeS spectra. Before comparing them, we combined the WiFeS's spectra into binary spectra by shifting them with the radial velocities of the primary and secondary components of CU Cnc obtained in Sect.\,\ref{sec:orbit} and then used the formula
\begin{equation}
    S_{\mathrm{SB}}=\frac{S_{\mathrm{A}}}{1+1/LR}+\frac{S_{\mathrm{B}}}{1+LR}.
\end{equation}
Here $LR$ is the luminosity ratio ($L_{\mathrm{A}}/L_{\mathrm{B}}$) and $S_\mathrm{A}$, $S_\mathrm{B}$, and $S_\mathrm{SB}$ correspond to the primary, secondary and combined spectra respectively. We adopt $LR=1.3$, \cla{which is the same as the value obtained from spectroscopy in Sect. \ref{sec:SS} and is close} to the values reported by \citetalias{ribas:2003} and \citetalias{wilson:2017}. As the components of CU Cnc are relatively similar, we assume that the binary spectra can be constructed by combining two identical template spectra corresponding to the same stellar parameters (i.e. $S_\mathrm{B}=S_\mathrm{A}$). We then performed a cross-correlation in order to see which stellar parameters of the WiFeS sample correlated the best with the spectra of CU Cnc. This analysis was carried out for all 20 Stokes $I$ spectra in the spectropolarimetric time series. The combined cross-correlation for all observations, normalised to the highest value, can be seen in Fig.~\ref{fig:ccf}.

We find that this analysis favours a super-solar metallicity for CU Cnc. The obtained $T_\mathrm{eff}$ is $3200\pm200$ and [Fe/H] is $0.2\pm0.2$, the error bars are selected based on the region where all stars have a cross-correlation peak within 5\% of the highest cross-correlation value. While this metallicity is on the lower end compared to the results of \citetalias{delfosse:1999} and \citetalias{wilson:2017}, it does support the super-solar \cla{[Fe/H] between 0.2 and 0.4 considered} by \citetalias{feiden:2013} for the stellar evolution modelling of CU Cnc. 
While a precise metallicity value is difficult to assign, it does appear that the Castor argument by \citetalias{ribas:2003} is not valid for CU Cnc as several different methods have arrived at a super-solar metallicity.

\subsection{Presence of lithium}
\citetalias{ribas:2003} found tentative traces of lithium on both components of CU Cnc using the 6707\,\AA\, lithium feature. They reported an equivalent width of $\sim50$\,m\AA\ for both components, which corresponded to a lithium abundance of $\log N_\mathrm{Li}/N_\mathrm{tot}\sim-13$. 
Lithium is expected to be depleted in M-dwarfs due to the fact that convective motions transport lithium deep into the stellar interior where the temperature is sufficiently high to destroy lithium. The presence of lithium absorption can be used to obtain an age estimate for young M-dwarfs. It is possible that magnetic fields could mitigate the lithium depletion as \cite{barradoynavascues:1997} found a correlation between activity and lithium abundance. This would be due to convective inhibition that reduces the rate at which lithium depletion takes place due to the fact that convection will be unable to transport lithium to the same depth within the star.

However, the existence of lithium on the surface of the components of CU Cnc has not been confirmed. Investigation by \citetalias{wilson:2017} found no traces of lithium in their observed spectra, although they still placed an upper limit of 50\,m\AA\ on the equivalent width. Furthermore, attempts by \cite{MacDonald:2015} and \citetalias{feiden:2013} to produce magnetoconvective models of CU Cnc have failed to find the reported lithium abundance of CU Cnc while simultaneously being consistent with other stellar parameters. 
As the spectra obtained with ESPaDOnS have both higher resolution and S/N compared to those used by \cite{wilson:2017}, 
it is worthwhile to revisit the claimed lithium detection.

Investigating the observed spectra, we find no obvious features corresponding to the \ion{Li}{i} 6707~\AA\ line at the expected wavelengths for either component. The lithium feature is hidden within a TiO band, making it challenging to identify any individual feature as lithium. In addition, by comparing synthetic spectra generated with the \textsc{Synmast} code \citep{Kochukhov:2007,kochukhov:2010} and \textsc{MARCS} \citep{gustafsson:2008} model atmospheres with different lithium abundances we find that the spectrum with similar abundance as reported in \citetalias{ribas:2003} would be essentially indistinguishable from the spectrum without any lithium. While this result does not change the upper limits as reported by \citetalias{ribas:2003}, the risk of including some equivalent width from nearby TiO lines makes it likely that this upper limit is an overestimation, especially since a complete absence of lithium absorption is not inconsistent when comparing observations with synthetic spectra.

\section{Large-scale magnetic field structure}
\label{sec:LS}

In order to obtain the large-scale magnetic field structures on stellar surfaces the use of Stokes $V$ spectra is required. While singular observations can give some insights into the field properties by measuring the longitudinal magnetic field strengths as described in Sect.~\ref{sec:FAP}, more information can be obtained by observing a time-series of the star as it rotates. This allows one to utilise the ZDI technique by combining information from different rotational phases in order to construct a surface distribution of the magnetic field vector. 

In order to compare with observations, we compute synthetic Stokes profiles assuming a line with a  Zeeman triplet splitting with the average line parameters of the LSD line mask from Sect.~\ref{sec:LSD}. The line profile is calculated using the Milne-Eddington approximation \citep[see][]{landi-deglinnocenti:2004}. The validity of single line interpretation of LSD profiles has been investigated by \cite{kochukhov:2010}. They found that the assumption is valid for circularly polarised spectra not exceeding 2\,kG. It is rare for cool stars to have large-scale field structures reaching these values. 
CU Cnc appears to be no exception, as the longitudinal field strengths of the components of CU Cnc presented in Table~\ref{tab:fap} do indicate field strengths significantly below this limit. For this reason these assumptions should not lead to significant shortcomings in interpretation of our LSD profiles.

We use the \textsc{InversLSDB} code presented in \cite{rosen:2018} to simultaneously obtain surface maps of the binary components of CU Cnc. This inversion code is capable of describing the surfaces of the binary components either using Roche lobe equipotentials, corresponding to the situation when components are tidally locked and co-rotating, or as spherical bodies rotating with independent rotation rates, possibly including differential rotation. We opt for the co-rotating Roche lobe geometry as the system is a close binary and previous studies suggested synchronisation of the rotation and orbital motion. The Roche lobe surface potentials are determined by adjusting them until the radii of both components correspond to the literature values presented in Table~\ref{tab:Stellar_Param}. While this does allow for the stellar shape to deviate from spherical, we find no significant deviation from spherical geometry caused by the gravitational interaction of the two components (about 0.03 \% of the stellar radius for both components).
Another parameter that needs adjusting is the relative local brightness. This parameter determines the fraction of the brightness of two surface elements of equal area on the two components. This value was determined to be 1.42 by finding the best fit to the Stokes $I$ profiles, assuming a homogeneous surface.
The magnetic field is described by using spherical harmonic functions \citep[e.g.][]{kochukhov:2014}. By excluding the 0th degree, these functions have the property of ensuring a divergence-free field while simultaneously giving valuable information about the field complexity on the surface. The code is also capable of accounting for phase smearing by calculating multiple profiles during the span of each observation. This is important as the individual Stokes $V$ observations take up $\sim2\%$ of the total orbital period, resulting in a radial velocity shift of up to about 8\,km\,s$^{-1}$ due to the orbital motion. Each observed phase is modelled with 5 sub-phases that are then integrated to find the final Stokes profiles.

While the inversion method can reconstruct both the surface magnetic field and the surface brightness, here we do not  include the surface brightness distribution. Even if the simultaneous recovery of both surface properties has been shown to improve the magnetic field reconstruction \citep{rosen:2012}, the Stokes $I$ LSD profiles of CU Cnc showed little indication of distortions due to spots over different phases. This, in combination with  relatively low rotation rates of the components ($<10$\,km\,s$^{-1}$), indicates that a brightness map will have a low impact on the reconstructed magnetic field.

\begin{figure*}[!th]
    \centering    \includegraphics{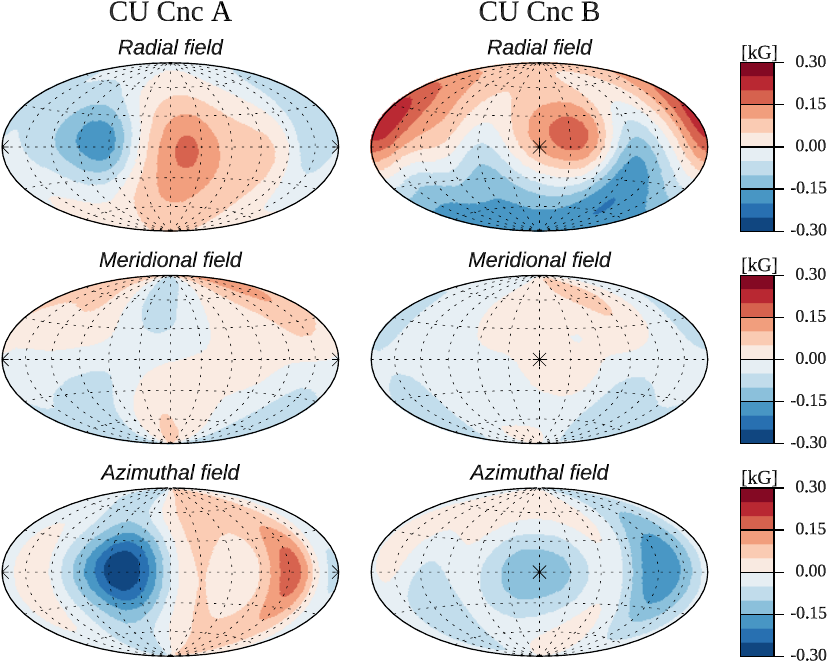}\includegraphics[height=0.45\textheight]{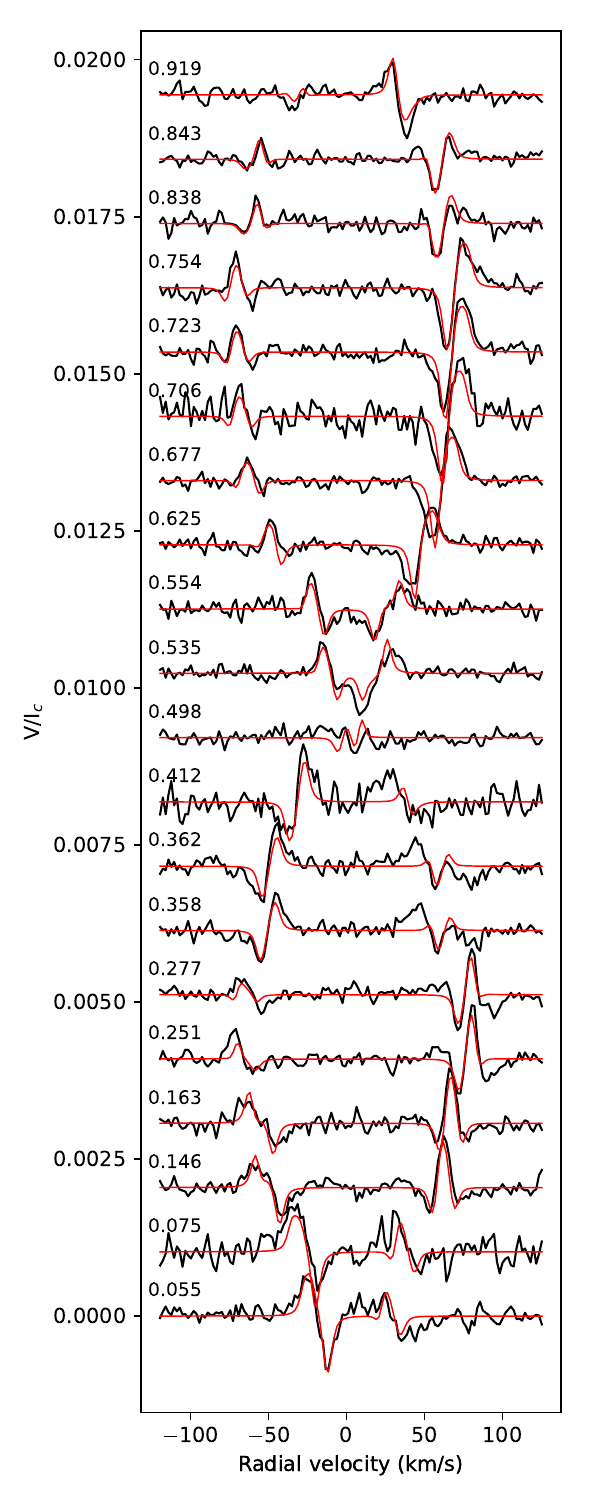}
    \caption{Results of the ZDI analysis of CU Cnc. \textit{Left:} Surface magnetic fields of CU Cnc shown in the Hammer-Aitoff projection in order to preserve the area of surface structures. The black asterisks represent the substellar point on each component. \textit{Right:} Observed Stokes $V$ LSD profiles  (black) and their synthetic counterparts (red) generated from the surface distributions on the left.}
    \label{fig:ZDI_map}
\end{figure*}

Another aspect that has been found to occasionally affect the reconstruction of magnetic fields on M-dwarfs is the use of a magnetic filling factor. This approach postulates that each surface element is only partially covered by the global magnetic field component. The reason for its inclusion is that \cite{morin:2008} found that the observed Stokes $V$ profiles are often too broad compared to regular synthetic profiles. The introduction of a filling factor allows the synthetic profile to become wider without simultaneously becoming too strong, improving the quality of the fit. Typical values used for the global-field filling factors are around 
10--15\,\% \citep[e.g.][]{morin:2008,morin:2010,lavail:2018,donati:2023}. We explored the impact of the choice of this parameter and found a relatively small dependence on the fit quality and resulting field structure. Only when adopting very low filling factors of around 10\,\% and less does the structure change significantly. 
Still, using a very low filling factor does not significantly improve the fit. In fact, we found the optimal magnetic filling factor to be around 20\,\%, which is the value we adopted for our investigation.

As ZDI is an ill-posed problem, this means that further constraints are required in order to obtain a unique and stable solution. This is done by minimising regularisation functions that penalise certain surface structures deemed to be too complex. In our implementation of ZDI, the magnetic field structure is regularised by the following penalty function \citep[e.g.][]{rosen:2018}
\begin{equation}
\label{eq:magreg}
    R_{B} = \Lambda_{B}\sum_{\ell=1}^{\ell_{\mathrm{max}}}\sum_{m=-\ell}^{\ell}\ell^2(\alpha_{\ell,m}^2+\beta_{\ell,m}^2+\gamma_{\ell,m}^2).
\end{equation}
This function penalises the degree $\ell$ of the magnetic field structure, where a higher degree corresponds to a more complex magnetic field distribution. It also penalises the strength of the magnetic field by favouring lower values of the spherical harmonic coefficients $\alpha$, $\beta$, and $\gamma$. $\ell_{\mathrm{max}}$ is the maximal spherical harmonic angular degree that is included in the inversion. From the rotational velocity of CU Cnc it is possible to determine the maximal angular degree that is resolvable following \cite{fares:2012}. For ESPaDOnS, the maximal degree that should be resolvable is $\approx9$. For this reason, we do not include any degree with $\ell>10$ in the inversion. This function will favour the simplest and weakest field topology that can fit observations.
In order to determine the value of $\Lambda_B$ to use for regularisation, we repeated ZDI inversions for multiple different regularisation strengths and compared the quality of the fit with the strength of the regularisation. We selected the regularisation at the point where reducing it further would give no significant improvement to the fit quality, which corresponded to $\Lambda_{B}=5\cdot10^{-10}$.

\begin{figure}
    \centering
    \includegraphics[width=\linewidth]{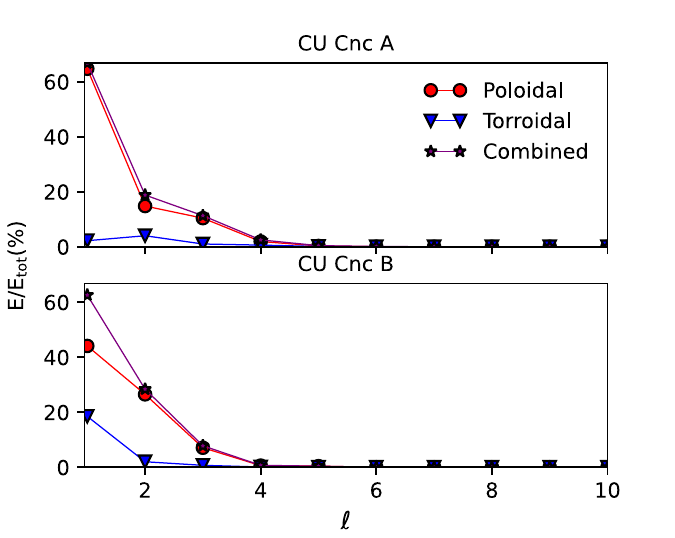}
    \caption{Magnetic field energy as a function of angular degree $\ell$. Both contributions from the poloidal (red circles) and toroidal (blue triangles) are shown, as well as the combination of the two (purple stars).}
    \label{fig:energydist}
\end{figure}

We performed the final ZDI analysis with the adopted parameters and obtained a surface map that can be seen in Fig.~\ref{fig:ZDI_map}. Our reduced $\chi^2$ for the Stokes V fit corresponding to the surface map is 1.26. 
\cla{In the phase interval between 0.35 and 0.42, we observed a feature in the Stokes $V$ profiles of CU Cnc B that is not present in our synthetic fit. This missing feature is not a consequence of chosen parameters, as its recovery is not dependent on regularization.}
This could indicate a change in field structure as the three observations were obtained within three days of each other towards the end of the observation sequence. Similarly to the longitudinal field measurements from Sect.~\ref{sec:FAP}, we find a surface distribution with surface fields on the order of $\sim100$\,G on both stars. The maximum field strength \cla{(after multiplying by the filling factor)} obtained 
on each component is about 330 and 240\,G \cla{for CU Cnc A and B, respectively}. Distribution of the magnetic energy over different harmonic modes is a common approach to characterise stellar magnetic fields \citep[e.g.][]{see:2015}. This includes the fraction of energy distributed in the poloidal and toroidal modes, the fraction of axisymmetric fields, and variation of energy with the angular degree. This assessment is presented in Table~\ref{tab:bprop} as well as in Fig.~\ref{fig:energydist}.

\begin{table}
    \centering
    \caption{Large-scale magnetic field parameters for CU Cnc.}
    \label{tab:bprop}
    \begin{tabular}{lrrrrr}
        \hline \hline
         & & \multicolumn{2}{c}{CU Cnc A} & \multicolumn{2}{c}{CU Cnc B} \\
        \hline
        \multicolumn{2}{l}{$\langle B_{V}\rangle$ (G)} & \multicolumn{2}{c}{117} & \multicolumn{2}{c}{128} \\
        \multicolumn{2}{l}{$|B_V|_{\mathrm{max}}$ (G)} & \multicolumn{2}{c}{329} & \multicolumn{2}{c}{243} \\
        \multicolumn{2}{l}{$E_{\mathrm{pol}}/E_{\mathrm{tot}}$(\%)} & \multicolumn{2}{c}{92.1} & \multicolumn{2}{c}{78.7} \\
        \multicolumn{2}{l}{$E_{m=0}/E_{\mathrm{tot}}$(\%)} & \multicolumn{2}{c}{8.9} & \multicolumn{2}{c}{66.5} \\
        \multicolumn{2}{l}{$E_{|m|<\ell/2}/E_{\mathrm{tot}}$ (\%)} & \multicolumn{2}{c}{10.7} & \multicolumn{2}{c}{66.8} \\
        \hline \hline
         & & $E_{\mathrm{pol},\ell}$ & $E_{\mathrm{tot},\ell}$ & $E_{\mathrm{pol},\ell}$ & $E_{\mathrm{tot},\ell}$ \\
        \hline
     $\ell=$ &  1  & 67.0 & 68.3 & 42.4 & 60.0 \\
        & 2 &  15.7 & 18.6 & 28.4 & 30.4\\
        & 3 &  9.4  & 10.3 & 7.5 & 8.3\\
        & 4 &  1.8  &  2.3 & 0.7 & 0.8\\
        & 5 &  0.3  &  0.4 & 0.4 & 0.4\\
        & 6 &  <0.1 &  $\sim$0.1 & 0.0 & 0.0\\
        \hline
    \end{tabular}
\end{table}

\section{Small-scale magnetic field}

\label{sec:SS}
In order to measure the small-scale fields, we rely on a group of magnetically sensitive \ion{Ti}{I} lines at slightly below $10000\,\AA$. These lines have been used in many recent studies of cool stars \citep[e.g.][]{shulyak:2019,hahlin:2022} as they have several beneficial properties for magnetic field determination. The lines belong to the same multiplet formed between atomic energy levels a$^5$F and z$^5$F$^{\rm o}$, meaning that there is no relative uncertainty in their line strength which reduces the impact of line parameter uncertainties when measuring magnetic field. Another advantage is that these lines have a range of magnetic field sensitivities, including a line with an effective Land\'e factor of zero. This means that this line has no response to the magnetic field, allowing magnetic and non-magnetic line broadening parameters to be constrained with less degeneracy. 

The \ion{Ti}{i} lines are studied using the disentangled spectra derived in Sect.~\ref{sec:orbit}. 
This means that we perform the analysis using time-averaged spectra, disregarding possible variation of small-scale fields across the stellar surfaces. While investigation of this variation would be interesting, the time-dependent blending between the two components would make this challenging, especially when telluric absorption also contributes to the spectra. In any case, studies looking into the rotational modulation of the small-scale fields such as \cite{bellotti:2023} found no strong indication of rotational modulation.

For the magnetic inference, we
generated a grid of synthetic spectra using stellar model atmospheres from \textsc{MARCS}, line lists from VALD, and the polarised radiative transfer code \textsc{Synmast}. We used a linear interpolation between the model atmosphere grid points to derive spectra corresponding to  $T_{\mathrm{eff}}$ and $\log g$ of CU Cnc components. The Ti abundance was allowed to vary. For the other elements, we assumed the solar abundance pattern from \cite{asplund:2009}. While previous stellar parameter analyses indicated a super-solar metallicity, we found that the magnetic field measurement is not sensitive to choice of metallicity. The only parameter significantly affected is the Ti abundance that correlates positively with metallicity. 

The spectra were calculated for the magnetic field in steps of 2\,kG, a typical value for cool star magnetic field investigations using optical and near-infrared high-resolution spectrographs \citep[e.g.][]{shulyak:2019,lavail:2019,petit:2021}. The magnetic field was assumed to be purely radial. While unlikely to be true given the large-scale field map illustrated in Fig.~\ref{fig:ZDI_map}, this standard assumption has been shown by \cite{kochukhov:2021} to not have a significant impact on the Stokes $I$ line shapes as the radial field configuration naturally produces a range of different field directions relative to the line of sight.

In order to account for the spread of field strength values, we utilised a multi-component model. Each magnetic field strength bin was assigned a filling factor $f_{i}$ corresponding to a fraction of the stellar surface covered by that magnetic field strength. The synthetic stellar spectrum $S$ is then given by
\begin{equation}
    S = \sum_{i}f_{i}S_{i},
\end{equation}
where $S_{i}$ are the synthetic spectra with specific field strength $B_{i}$. The mean field strength is calculated using $\langle B_I \rangle=\sum f_i B_i$.

In order to find the optimal parameters and their uncertainties we used the SoBAT library for IDL \citep{anfinogentov:2021} to carry out Markov chain Monte-Carlo (MCMC) sampling. In our approach, described by \cite{hahlin:2022}, we obtain the small-scale magnetic field parameters on both binary components fitting their spectra simultaneously. 
\cla{Besides the magnetic filling factors, the free parameters include the $v\sin i$ of the components, their shared Ti abundance and the luminosity ratio $LR$.}
As discussed by \cite{hahlin:2022}, spectral disentanglement produces spectra with an arbitrary radial velocity zero point. For this reason, radial velocity is also included as a free parameter for each component. For M-dwarfs it is difficult to accurately normalise the spectra. Consequently, we also include continuum scaling for each line as a free parameter following \cite{shulyak:2019}. 
In an analysis of high S/N, high resolution observations, systematic biases often dominate over random observational errors, which means that only accounting for the observational errors typically underestimates uncertainties. SoBAT optionally allows for the spectral variance to be treated as a free parameter. We use this technique here to obtain more realistic uncertainties. All parameters are given uniform priors and best-fitting parameter values are assumed to be equal to the median of the posterior distributions. To avoid non-physical solutions, an additional constraint is introduced that requires the sum of the filling factors to not exceed 1.

In principle, one could add an arbitrary number of magnetic field filling factors to the model. This is however problematic, as increasing the number of filling factors also tends to increase the field strength as pointed out by e.g. \cite{shulyak:2019} and \cite{petit:2021}. In order to avoid strong spurious fields that might skew our results towards stronger average magnetic field strengths we use the Bayesian information criterion \citep[BIC]{sharma:2017} to penalise models using more free parameters to describe the magnetic field. We do this by iteratively adding stronger magnetic field components to the model until no significant improvement to the fit is obtained. To ensure that the MCMC method finds the optimal region before mapping posterior distributions, we use a burn-in length of 20000 steps. We then run the MCMC sampling until a sufficient number of independent samples have been collected. This is quantified by the effective sample size (ESS) determined by the autocorrelation time. Our threshold is an ESS of 1000. 

\begin{table}
    \centering
    \caption{Small-scale magnetic field parameters for CU Cnc.}
    \label{tab:small_scale}
    \begin{tabular}{lcc}
        \hline \hline
         & CU Cnc A & CU Cnc B \\
        \hline
         $f_2$& $0.676\pm\substack{{0.031}\\{0.040}}$& $0.542\pm\substack{{0.030}\\{0.038}}$ \\
         $f_4$& $0.064\pm\substack{{0.053}\\{0.041}}$& $0.059\pm\substack{{0.056}\\{0.039}}$ \\
         $f_6$& $0.241\pm\substack{{0.022}\\{0.027}}$& $0.380\pm\substack{{0.022}\\{0.027}}$ \\
         $v\sin i$ (km\,s$^{-1}$) & $9.15\pm\substack{0.18\\0.19}$ & $8.44\pm\substack{0.24\\0.23}$ \\
         $\varepsilon_{\ion{Ti}{I}}$& \multicolumn{2}{c}{$-7.76\pm0.01$} \\
         $LR$ & \multicolumn{2}{c}{$1.30\pm0.01$}\\
         \hline
    \end{tabular}
    \tablefoot{Values are presented with their 68\% confidence regions.}
\end{table}
\begin{figure*}
    \centering
    \includegraphics[width=\textwidth]{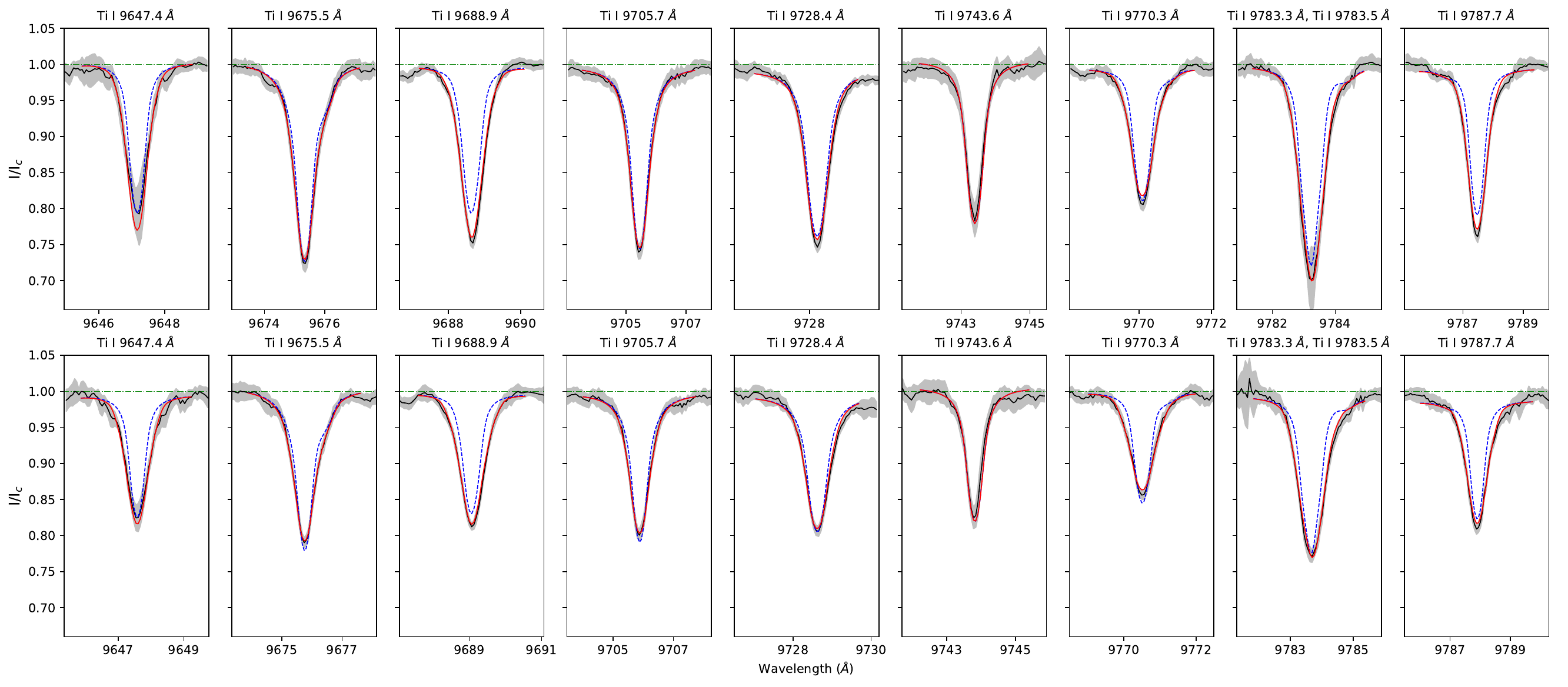}
    \includegraphics[width=0.45\textwidth]{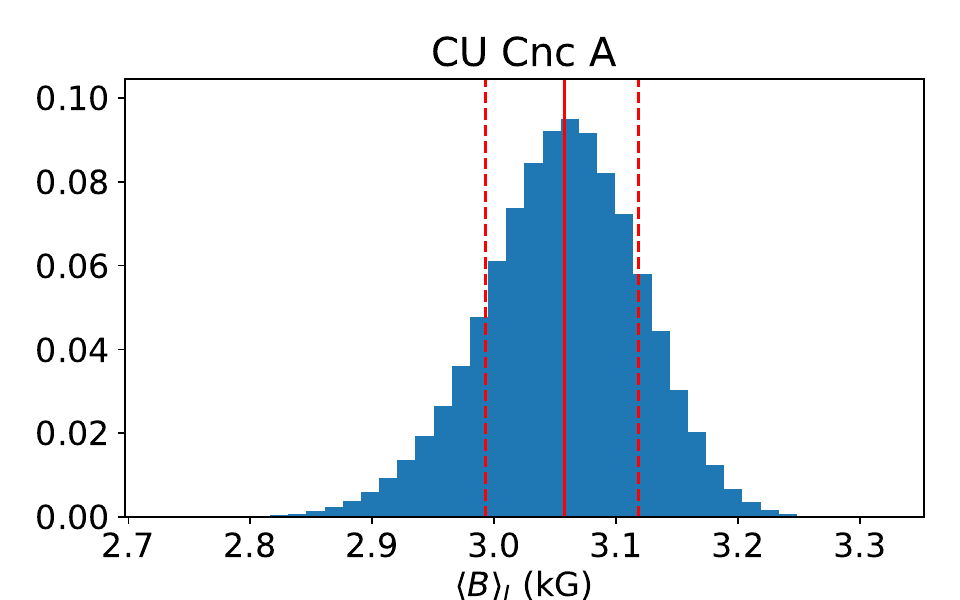}
    \includegraphics[width=0.45\textwidth]{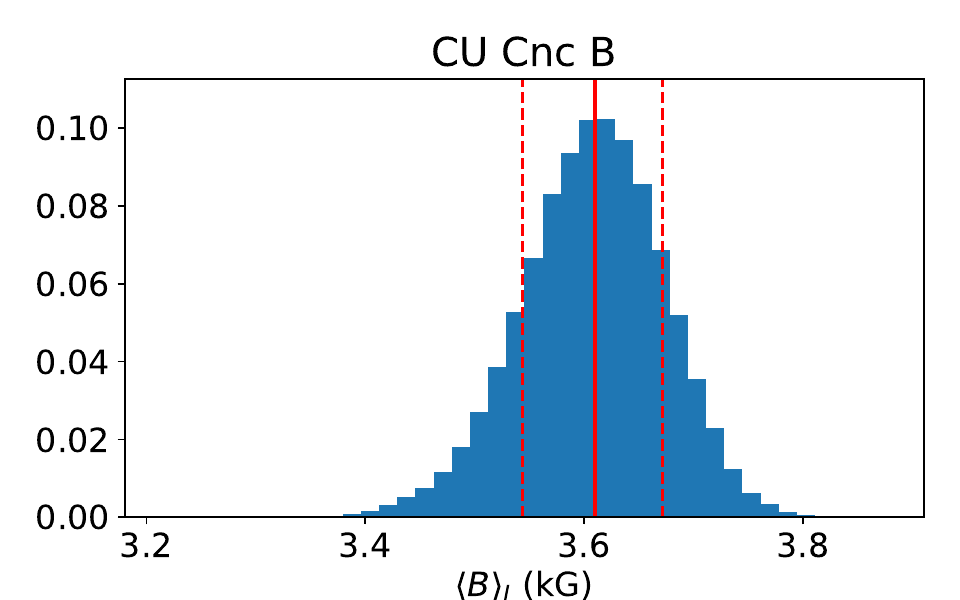}
    \caption{Fits to the disentangled spectra and resulting posteriors of the average small-scale magnetic field strength. \textit{Top:} The studied \ion{Ti}{i} lines, with observations in black, median parameter model in red, and non-magnetic spectra with otherwise identical parameters in dashed blue. The two rows correspond to the primary (first row) and secondary (second row) components. \textit{Bottom:} Posterior distributions of the average magnetic field strength of the two components with the median and 68\% credence regions marked with vertical red lines.}
    \label{fig:small_scale_results}
\end{figure*}

We found that a model including of 4 components (one zero-field and three magnetic) was the most suitable to describe the small-scale surface fields of CU Cnc. When using this model, we obtained average small-scale magnetic fields with strengths of 3.1--3.6 kG for both components of CU Cnc. The average field strengths are reported in Table~\ref{tab:results}. Using another number of filling factors has a small, but statistically significant, influence on the result. For example, adding or removing one component from the MCMC sampling shifts the overall field strength by $\lesssim0.2$\,kG for the components. The field has a strong influence on the line profiles, as seen in Fig.~\ref{fig:small_scale_results}, showing both intensification and broadening. The filling factors in Table~\ref{tab:small_scale} show that essentially the entire surface is covered in $\sim$kG magnetic fields for both components. It also appears as if the field is split into two groups, as it is primarily the weakest and strongest component that contribute to the spectra. 
While this dichotomy could be due to the magnetic field model, as the 4\,kG filling factor is highly correlated with the other two (see Fig~\ref{fig:small_scale_corner}), it could also mean that the magnetic field on the surface consist of two different structures with different magnetic properties.

We found a luminosity ratio of 1.3, which is in close agreement with the results by \citetalias{ribas:2003} and \citetalias{wilson:2017}. The obtained titanium abundance is  significantly lower than what would be anticipated from the super-solar metallicity of CU Cnc. The likely cause for this is that the equivalent width of \ion{Ti}{I} lines are strongly anti-correlated with overall metallicity due to formation of the TiO molecule. By only changing the titanium abundance this trend is not recovered. In any case, the role of the Ti abundance in the context of this investigation is to set a base line depth for each line in order to derive magnetic parameters. Our $v\sin i$ values \cla{of 9.15 and 8.44\,km\,s$^{-1}$} are slightly larger than the 7.9 and 7.1\,km\,s$^{-1}$ that would be predicted from a tidally locked system with the parameters given in Table~\ref{tab:Stellar_Param}.
This is likely due to the fact that we have not included a macroturbulent broadening in the synthetic spectrum generation.

\section{Discussion}
\label{sec:discussion}

\begin{table}
    \centering
    \caption{Observed and theoretical magnetic field parameters of CU Cnc.}
    \label{tab:results}
    \begin{tabular}{lcc}
        \hline\hline
         & CU Cnc A & CU Cnc B \\
        \hline
        $\langle B_{V}\rangle$ (G) & 117 & 128 \\
        $\langle B_{I}\rangle$ (kG)& $3.06\pm0.06$ & $3.61\pm0.06$ \\
        $\langle B_{V}\rangle/\langle B_{I}\rangle$ (\%) & 3.8 & 3.5 \\
        \citetalias{feiden:2013}$^*$ (kG) & \multicolumn{2}{c}{2.6-3.5}\\
        \citetalias{macdonald:2014}$^*$ (kG) & \multicolumn{2}{c}{0.45-0.52}\\
        \hline
    \end{tabular}
    \tablefoot{A $^*$ Represents range of values providing a satisfactory radius fit given by the references.}
\end{table}

\subsection{Comparison between different spatial scales}
\label{subsec:spatial}
The difference between magnetic field strengths obtained from small- and large-scale measurements is well established for M-dwarfs \citep{kochukhov:2019,see:2019}. Our result does not contradict this as the ratio between the obtained average field strengths shown in Table~\ref{tab:results} reveals that less than 5\,\% of the small-scale field strength is recovered on large spatial scales. \cite{kochukhov:2019} also showed that the recovery fraction is dependent both on the strength of the small-scale field, the complexity, and the axisymmetry of the large-scale field. Both of the strength and axisymmetry tends to increase the recovery fraction while complexity tends to decrease it. In the context of this, CU Cnc appears to be on the low end of the large-scale magnetic strength recovery. 

This low recovery fraction appears primarily to be due to a rather weak and complex large-scale field as compared to other M-dwarfs slightly above the fully convective limit investigated by \cite{morin:2008}. The small-scale field strengths are similar to what has been reported by e.g. \cite{shulyak:2017}, which indicates that our small-scale results should not systematically reduce the recovery fraction. In fact, as we mentioned in Sect.\,\ref{sec:SS} we elect to stop adding filling factors beyond 6\,kG, which is earlier than in \cite{shulyak:2019}. 
This has an effect of somewhat reducing $\langle B_I\rangle$, \cla{which would result in an increased recovery fraction for CU Cnc compared to the recovery fractions of the stars discussed by \cite{kochukhov:2019} since many of those results relied on the analysis by \cite{shulyak:2017} who used filling factors corresponding to field strengths up to 10\,kG.}
While the low recovery fraction is likely real, it is possible that it is also significantly reduced by the hemisphere degeneracy explored in \cite{hahlin:2021} and visible in Fig.~\ref{fig:ZDI_map}. Using ZDI on high inclination targets has a tendency to significantly underestimate contribution of the spherical harmonic modes that are antisymmetric with respect to the equator, which also reduces the measured $\langle B_V\rangle$.

\subsection{Comparison between components}
\label{subsec:components}
As the non-magnetic properties of the CU Cnc components are relatively similar to each other it is also interesting to compare the obtained magnetic properties between the two components. It seems as if the secondary component has a stronger magnetic field on both spatial scales, although only slightly on the large scales\cla{. With a difference of about $9\sigma$ in the small-scale fields, CU Cnc B has a significantly stronger total field.}
This difference in the overall field strength could originate from the lower mass of the secondary, which will result in a \cla{longer convective turnover time. Using the empirical relation between convective turnover time and mass from \cite{wright:2011}, we find values of 42.4 and 46.5 for CU Cnc A and B, respectively}. Since the stars are tidally locked and have the same rotational period, this means that the Rossby number \cla{is about 10\,\% lower for the secondary}. As magnetic field strengths tend to increase with decreasing Rossby number \citep[e.g.][]{see:2015}\cla{, this could result in a stronger field on the secondary}. 
It is however problematic that the components \cla{are} in the saturated regime where the trend with Rossby number should be much less significant. Comparing with the results from \citet{reiners:2022}, it also appears that the difference obtained in this study is well within the scatter in the saturated regime. This could indicate that there is some other variation that causes the difference. One such possibility could be similar to what \cite{bellotti:2023} found when monitoring the small-scale field evolution of AD Leo, a star with similar mass and rotation period ($0.42\,M_\odot$ and $2.23$\,days) as CU Cnc, and found a trend that seems to correlate with the stellar activity cycle. The measured small-scale fields of the CU Cnc components falls close the range of reported values for AD leo (2.8-3.6\,kG) during this evolution. This indicates that the difference seen for the components of CU Cnc could be due to some small-scale field evolution coupled to an activity cycle that is out of sync between the two components.

\cla{When considering the H$\alpha$ emission of CU Cnc, it appears that CU Cnc A exhibits a systematically stronger emission. According to \cite{reiners:2022}, the H$\alpha$ luminosity should correlate with the magnetic flux. When comparing the magnetic fluxes of the two components using the field strength measurements from Table~\ref{tab:results} and the radii from \citetalias{ribas:2003}, we find very similar magnetic fluxes (about 2\% difference). This means that the strength of the intrinsic H$\alpha$ emission should be similar for the two components while we found emission to be slightly stronger for the primary. At the same time, considering the scatter in the magnetic field-H$\alpha$ flux relation in \cite{reiners:2022}, the difference between the H$\alpha$ emission of the two components obtained in Sect.~\ref{sec:emission} does not stand out as particularly significant. Furthermore, as CU Cnc B is fainter of the two stars, its emission in the composite spectrum is reduced to a greater extent due to the continuum dilution. Assuming the same intrinsic H$\alpha$ equivalent widths and luminosity ratio of 1.3, the secondary should exhibit 23\% weaker H$\alpha$ in the composite spectrum compared to 16\% weaker emission reported in Sect.~\ref{sec:emission}. 
Thus, the observed relative H$\alpha$ emission and magnetic field strength of the two stars are not inconsistent with each other.
}

In some respects, the structure of the obtained large scale field also differs quite significantly. 
Even if the average field is marginally stronger on the secondary, the primary exhibits a significantly stronger peak strength.
While both stars exhibit a predominantly poloidal field structure, the secondary component has a mostly axisymmetric field geometry while only about 10\% of the field energy is axisymmetric for the primary. The secondary component has a significant toroidal structure accounting for about 20\,\% of the total energy. Single stars close to the convective boundary are almost entirely dominated by poloidal structures and mostly axisymmetric \citep[e.g.][]{morin:2008,bellotti:2023}. This means that the magnetic structure of both components of CU Cnc are unusual in different ways, indicating that CU Cnc a system with unique magnetic properties.

One thing to note about the surface structures before making any conclusions, is that the magnetic structure of the primary component shown in Fig.~\ref{fig:ZDI_map} exhibits symmetric structures with respect to the equator. This could be due to hemisphere degeneracy of high inclination targets caused by large-scale cancellation of opposite field polarities as discussed in \cite{hahlin:2021}. The effect of this is that the observed polarisation profiles contain
information only from a subset of spherical harmonic modes. The reason that this effect is less pronounced on the secondary component is due to the fact that eclipses can help to mitigate the degeneracy \citep[e.g.][]{vincent:1993}. In this case, the Stokes $V$ observation around phase 0.498 is a partial eclipse where the primary blocks the secondary, which would help to mitigate some of the hemisphere degeneracy on the secondary component. 
This effect is likely partially responsible for the weak axisymmetric contribution on the primary component, as signal from a dipole aligned with the rotation axis would be mostly cancelled out due to the high inclination. To verify that this is indeed the case, linear polarisation observations of CU Cnc would need to be successfully carried out.

The overall complexity of the magnetic fields on the two components is relatively similar. Fig.~\ref{fig:energydist} shows that both stars are dominated by dipole structures with some significant contributions from the quadru- and octupole. Only about 0.1\% of the magnetic energy is contained within structures with complexities beyond $\ell=6$ for either component, which justifies our $\ell_\mathrm{max}=10$ cutoff.

We can also compare the location of the substellar point shown in Fig.~\ref{fig:ZDI_map} with the magnetic structures. What we find for the primary is that the substellar point is not connected to any particular structure on the stellar surface. For the secondary the result is slightly different. While the substellar point is not close to the strongest magnetic feature, it does coincide with one of the stronger regions on the surface. While this could be coincidental, especially since the stars are quite separated, it is possible that the magnetic spot on the secondary arises due to the interaction with the primary. To verify this one would need to follow the evolution of the surface field to ensure that this magnetic feature is stable over longer timescales compared to other magnetic features. 

\subsection{Comparison with theoretical modelling}
\label{subsec:theory}

The magnetic field of CU Cnc has been studied theoretically in order to investigate its influence on stellar evolution. To this end, \citetalias{feiden:2013} and \citetalias{macdonald:2014} have made predictions on magnetic field strengths from magnetoconvective stellar evolution models. These predictions, along with the measured values, can be seen in Table \ref{tab:results}. \citetalias{feiden:2013} found surface field strengths around 3\,kG for both components while \citetalias{macdonald:2014} required a much lower surface field strengths of around 0.5\,kG. 

Comparing our obtained values we find that our large-scale field strengths are too weak compared to both model predictions. The observed small-scale fields are, however, within the range of predictions made by \citetalias{feiden:2013}. Based on this good agreement with \citetalias{feiden:2013}, we can make estimates of the range of possible ages of the system using their evolutionary models. As the secondary component has a measured field strength value close to the presented 3.5\,kG model, CU Cnc would have an age between 0.3 to 6\,Gy. This range is narrower when also considering the intermediate field strength of $\sim3$\,kG on the primary, which would place the lower limit of the age of CU Cnc at slightly more than 1\,Gy. One aspect to consider is that the magnetic models calculated in \citetalias{feiden:2013} had a fixed metallicity of [Fe/H]$= +0.2$. Given the metallicity results from this work, \citetalias{delfosse:1999}, and \citetalias{wilson:2017}, its is possible that CU Cnc is more metal rich than this, which would shift the stellar radii somewhat. Regardless, the metallicity is a marginal effect and it is unlikely that changing this parameter would lower the range of possible ages to that of the Castor moving group, which was the claimed age by \citetalias{ribas:2003}.

This analysis of CU Cnc, in combination with the previous results on the eclipsing binaries YY Gem and UV Psc \citep{kochukhov:2019,hahlin:2021}, indicates that the magnetic treatment used in the models by \cite{feiden:2012a} is in a better agreement with the surface magnetic fields observed using high-resolution spectroscopy compared to the prescription used by \citetalias{macdonald:2014}.

\section{Conclusions}
\label{sec:conclusions}

In this work, the magnetic properties of the eclipsing binary CU Cnc have been investigated. We have characterised the magnetic fields on both large- and small scales using information from both the intensity and polarisation of stellar spectra. We have also compared our results with theoretical predictions from magnetoconvective stellar evolution models.

We found that the small-scale fields are about one order of magnitude stronger than magnetic structures on the large scale. This is in line with other investigations of M-dwarfs \citep[e.g.][]{kochukhov:2019,see:2019,kochukhov:2021}. It also appears that the inferred small-scale field strength agrees well with the theoretical predictions made by \citetalias{feiden:2013}. As this agreement has been seen for other stars as well \citep{kochukhov:2019,hahlin:2021}, it shows that their approach is reliable at linking the radius inflation to the surface field strength of low-mass stars.
While the work from \citetalias{macdonald:2014} present significantly weaker field strengths compared to our results, \cite{macdonald:2018} have been able to produce kG level field strengths when trying to replicate observational results from \cite{kochukhov:2017}. This shows that both models could reproduce the kG fields commonly observed and therefore benefit from being constrained by observational results. Furthermore, observations could also provide insight into which assumptions and parameters of the model are able to produce realistic field parameters which could help our understanding of the physics of stars. 

Using magnetic field measurements on other stars with inflated radii could therefore be a good way to both mitigate the radius discrepancy while simultaneously constraining stellar ages. This discussion also gives some indication of possible age of the CU Cnc system. It is likely that this binary is at least $\sim1$\,Gy old. This puts the previous tentative detection of lithium by \citetalias{ribas:2003} to question as it would be very difficult to prevent its destruction at this age. Especially since we could not confirm the detection of Li with the observational data analysed in this study. 

The large-scale magnetic field investigation of CU Cnc likely suffers from systematic errors in the form of hemisphere degeneracy due to high inclination. We see that this issue can be somewhat mitigated by eclipses. It might be advisable for future studies of eclipsing binaries to time a few observations in the time series such that they coincide with the eclipses of both components. As the orbital motion is typically well known for these systems, this should pose no major challenge as similar eclipse timings are regularly performed for exoplanet atmosphere transit studies. Another solution would be to obtain linear polarisation observations and include these data in future ZDI studies. Besides the improved smaller detail recovery in the ZDI inversions \citep[e.g.][]{rosen:2015}, \cite{hahlin:2021} showed that for high inclination targets linear polarisation carries more easily detectable information about the field components symmetric with respect to the equator. Due to the weak signal, observing linear polarisation is significantly more challenging as most instruments used for magnetic field investigations are not able to reach sufficient S/N to reliably detect the linear polarisation signal. The best candidate is likely the PEPSI spectrograph \citep{strassmeier:2018} at the LBT, as recent studies have shown the capability of PEPSI to recover Stokes $V$ LSD profiles at a very high S/N \citep[e.g][]{strassmeier:2023,metcalfe:2023}. Even if a time series observation of linear polarisation is unfeasible, \cite{kochukhov:2020b} showed that for AU Mic even a few observations of linear polarisation could be used to constrain a strong axisymmetric dipole field that was not visible from the Stokes $V$ profiles, giving valuable additional information on the magnetic field properties. Obtaining such observations on CU Cnc would be very useful in verifying that the large scale field structures obtained here are accurate. This is particularly interesting in this case as both components of CU Cnc show unusual field structures compared to the magnetic fields of other stars just above the convective limit \citep[e.g.][]{morin:2008,bellotti:2023}

\begin{acknowledgements}
Based on observations obtained at the Canada–France– Hawaii Telescope (CFHT) which is operated from the summit of Maunakea by the National Research Council of Canada, the institut National des Sciences de l’Univers of the Centre National de la Recherche Scientifique of France, and the University of Hawaii. The observations at the Canada–France–Hawaii Telescope were performed with care and respect from the summit of Maunakea which is a significant cultural and historic site.

A.H. and O.K. are acknowledging support by the Swedish Research Council (projects 2019-03548 and 2023-03667).

A.D.R acknowledges support by the KAW foundation (grant 2018.0192).

Software: NumPy \citep{harris:2020}, Matplotlib \citep{Hunter:2007}, Astropy \citep{AstropyCollaboration:2013}, iPython \citep{perez:2007}, Pandas \citep{mckinney:2010}, SciPy \citep{virtanen:2020}
\end{acknowledgements}

%\bibliographystyle{aa}
%\bibliography{references}

%APPENDIX
\onecolumn
\begin{appendix}

\section{Radial velocity measurements}
\begin{table*}[h!]
    \centering
    \caption{Radial velocities obtained from the Stokes $I$ LSD profiles.}
    \begin{tabular}{rrrr|rrrr}
    \hline\hline
    HJD & $V_A$ (km\,s$^{-1}$) & $V_B$ (km\,s$^{-1}$) & S/N$_\mathrm{I}$ & HJD & $V_A$ (km\,s$^{-1}$) & $V_B$ (km\,s$^{-1}$) & S/N$_\mathrm{I}$\\
    \hline
2455930.8874&71.6$\pm$0.3&-71.5$\pm$0.3&138&2455937.8785&-64.5$\pm$0.3&77.8$\pm$0.3&138\\
2455930.9011&71.5$\pm$0.3&-71.5$\pm$0.3&136&2455937.8922&-64.1$\pm$0.3&77.4$\pm$0.3&139\\
2455930.9147&71.5$\pm$0.3&-71.5$\pm$0.3&134&2455937.9059&-63.8$\pm$0.3&77.0$\pm$0.3&140\\
2455930.9284&71.3$\pm$0.3&-71.2$\pm$0.3&138&2455937.9196&-63.5$\pm$0.3&76.5$\pm$0.3&137\\
2455931.1187&63.2$\pm$0.3&-61.1$\pm$0.3&126&2455938.1039&-51.4$\pm$0.3&63.3$\pm$0.3&129\\
2455931.1324&62.1$\pm$0.3&-60.1$\pm$0.3&118&2455938.1175&-50.2$\pm$0.3&61.9$\pm$0.3&134\\
2455931.1461&61.0$\pm$0.3&-58.9$\pm$0.3&114&2455938.1312&-48.9$\pm$0.3&60.6$\pm$0.3&139\\
2455931.1597&59.8$\pm$0.3&-58.0$\pm$0.3&118&2455938.1449&-47.6$\pm$0.3&59.4$\pm$0.3&133\\
2455931.9738&-49.0$\pm$0.3&60.7$\pm$0.3&142&2455938.9866&63.2$\pm$0.4&-61.2$\pm$0.4&136\\
2455931.9875&-50.3$\pm$0.3&62.1$\pm$0.3&139&2455939.0002&64.1$\pm$0.4&-62.4$\pm$0.4&140\\
2455932.0012&-51.4$\pm$0.3&63.6$\pm$0.3&136&2455939.0139&65.0$\pm$0.4&-63.6$\pm$0.4&142\\
2455932.0149&-52.5$\pm$0.3&65.1$\pm$0.3&134&2455939.0276&65.8$\pm$0.4&-65.0$\pm$0.4&139\\
2455932.9483& & &137& 2455939.1160&70.0$\pm$0.3&-70.1$\pm$0.3&129\\
2455932.9619& & &138&2455939.1297&70.4$\pm$0.3&-70.5$\pm$0.3&133\\
2455932.9756& & &134&2455939.1434&70.8$\pm$0.3&-70.8$\pm$0.3&128\\
2455932.9893& & &136&2455939.1571&71.1$\pm$0.3&-71.0$\pm$0.3&120\\
2455933.1026&23.3$\pm$0.3&-17.9$\pm$0.3&119&2455940.8860&-50.6$\pm$0.3&62.0$\pm$0.3&124\\
2455933.1163&25.3$\pm$0.3&-19.9$\pm$0.3&128&2455940.8997&-49.3$\pm$0.3&60.6$\pm$0.3&127\\
2455933.1300&27.2$\pm$0.3&-22.0$\pm$0.3&126&2455940.9133&-48.0$\pm$0.3&59.4$\pm$0.3&132\\
2455933.1437&29.2$\pm$0.3&-24.4$\pm$0.3&132&2455940.9270&-46.5$\pm$0.3&58.0$\pm$0.3&131\\
2455933.9041&62.1$\pm$0.3&-60.0$\pm$0.3&145&2455941.0238&-35.2$\pm$0.3&46.0$\pm$0.3&89\\
2455933.9178&61.0$\pm$0.3&-58.9$\pm$0.3&145&2455941.0375&-33.0$\pm$0.4&43.5$\pm$0.4&69\\
2455933.9315&59.8$\pm$0.3&-57.8$\pm$0.3&139&2455941.0512&-31.6$\pm$0.4&41.5$\pm$0.4&93\\
2455933.9452&58.5$\pm$0.3&-56.7$\pm$0.3&138&2455941.0648&-30.1$\pm$0.4&39.9$\pm$0.4&46\\
2455934.1140&39.6$\pm$0.3&-36.0$\pm$0.3&128&2455941.8389&68.0$\pm$0.4&-68.0$\pm$0.4&88\\
2455934.1277&37.5$\pm$0.3&-34.2$\pm$0.3&130&2455941.8526&68.6$\pm$0.4&-68.5$\pm$0.4&70\\
2455934.1414&35.6$\pm$0.3&-32.3$\pm$0.3&134&2455941.8663&69.2$\pm$0.4&-69.4$\pm$0.4&63\\
2455934.1550&33.7$\pm$0.3&-30.3$\pm$0.3&136&2455941.8799&69.6$\pm$0.4&-69.9$\pm$0.4&51\\
2455935.0353&-65.0$\pm$0.3&78.3$\pm$0.3&141&2455942.8064&-16.5$\pm$0.3&25.7$\pm$0.3&110\\
2455935.0490&-65.0$\pm$0.3&78.3$\pm$0.3&142&2455942.8200&-18.5$\pm$0.3&27.9$\pm$0.3&116\\
2455935.0626&-65.0$\pm$0.3&78.3$\pm$0.3&141&2455942.8337&-20.4$\pm$0.4&30.0$\pm$0.4&91\\
2455935.0763&-64.9$\pm$0.3&78.3$\pm$0.3&143&2455942.8474&-22.8$\pm$0.3&32.1$\pm$0.3&85\\
2455935.8213&14.6$\pm$0.4&-9.1$\pm$0.4&145&2455942.8616&-24.8$\pm$0.3&33.9$\pm$0.3&122\\
2455935.8350&16.7$\pm$0.3&-11.8$\pm$0.3&144&2455942.8753&-26.7$\pm$0.3&35.6$\pm$0.3&131\\
2455935.8487&18.9$\pm$0.4&-14.1$\pm$0.4&143&2455942.8890&-28.6$\pm$0.3&37.5$\pm$0.3&131\\
2455935.8624&21.1$\pm$0.4&-16.0$\pm$0.4&142&2455942.9027&-30.1$\pm$0.4&39.8$\pm$0.4&43\\
2455936.0715&49.5$\pm$0.4&-46.8$\pm$0.4&141&2455943.1070&-53.1$\pm$0.3&65.5$\pm$0.3&130\\
2455936.0851&51.1$\pm$0.4&-48.9$\pm$0.4&140&2455943.1207&-54.1$\pm$0.3&66.9$\pm$0.3&133\\
2455936.0988&52.7$\pm$0.4&-50.8$\pm$0.4&143&2455943.1343&-55.2$\pm$0.3&68.2$\pm$0.3&132\\
2455936.1125&54.1$\pm$0.3&-52.4$\pm$0.3&143&2455943.1480&-56.2$\pm$0.3&69.3$\pm$0.3&119\\

\hline
    \end{tabular}
     \tablefoot{The last column lists the S/N of individual sub-exposures. Radial velocities are not given for the four observations around HJD=2455932.97. These observations were taken during an eclipse, which blended the lines of the components resulting in large measurement uncertainties. 
     }
    \label{tab:rv}
\end{table*}
\newpage
\section{Magnetic field measurements}
\begin{table*}[h!]
    \centering
    \caption{Magnetic measurements of individual Stokes $V$ profiles.}
    \label{tab:fap}
    \begin{tabular}{lcrrrcrcrr}
    \hline \hline
    Phase & HJD  & \multicolumn{2}{c}{$\langle B_z\rangle$} & \multicolumn{4}{c}{FAP} & S/N$_\mathrm{V}$ & S/N$_{\mathrm{LSD}}$  \\
     & ($\cla{+}24\mathrm{e}5$) & \cla{CU Cnc} A &  \cla{CU Cnc} B & \multicolumn{2}{c}{ \cla{CU Cnc} A} & \multicolumn{2}{c}{ \cla{CU Cnc} B} & & \\
    \hline
   0.055 & 55942.8269 & $115\pm24$ & $112\pm34$ & 0.000e+00 & (DD) & 2.923e-04 & (MD) & 208 & 7724\\
0.075 & 55942.8822 & $133\pm29$ & $129\pm46$ & 5.079e-07 & (DD) & 1.118e-02 & (ND) & 230 & 5093\\
0.146 & 55931.9944 & $91\pm23$ & $-2\pm13$ & 3.419e-14 & (DD) & 0.000e+00 & (DD) & 278 & 11476\\
0.163 & 55943.1275 & $92\pm25$ & $24\pm17$ & 6.826e-10 & (DD) & 0.000e+00 & (DD) & 263 & 10417\\
0.251 & 55935.0558 & $43\pm11$ & $35\pm16$ & 1.702e-08 & (DD) & 0.000e+00 & (DD) & 287 & 11623\\
0.277 & 55937.8990 & $34\pm10$ & $32\pm15$ & 4.355e-04 & (MD) & 0.000e+00 & (DD) & 279 & 11333\\
0.358 & 55938.1244 & $-80\pm19$ & $102\pm35$ & 0.000e+00 & (DD) & 3.162e-08 & (DD) & 267 & 10848\\
0.362 & 55940.9065 & $-95\pm22$ & $122\pm45$ & 0.000e+00 & (DD) & 2.825e-08 & (DD) & 258 & 10389\\
0.412 & 55941.0443 & $-115\pm32$ & $144\pm59$ & 6.780e-06 & (DD) & 9.429e-02 & (ND) & 153 & 4759\\
0.498$^*$ & 55932.9688 & $23.6\pm6.3$ & $19.6\pm5.6$ & 4.901e-03 & (ND) & 7.778e-03 & (ND) & 275 & 11293\\
0.535$^*$ & 55935.8418 & $-111\pm23$ & $95\pm32$ & 0.000e+00 & (DD) & 2.298e-14 & (DD) & 291 & 11657\\
0.554 & 55933.1232 & $-95\pm24$ & $52\pm25$ & 3.371e-10 & (DD) & 3.771e-09 & (DD) & 257 & 9918\\
0.625 & 55936.0920 & $-115\pm29$ & $-18\pm17$ & 0.000e+00 & (DD) & 5.141e-05 & (MD) & 287 & 11473\\
0.677 & 55939.0071 & $-155\pm40$ & $1\pm12$ & 0.000e+00 & (DD) & 1.934e-03 & (ND) & 282 & 11602\\
0.706 & 55941.8594 & $-128\pm29$ & $30\pm26$ & 3.677e-11 & (DD) & 2.426e-01 & (ND) & 142 & 4624\\
0.723 & 55939.1366 & $-134\pm26$ & $27\pm14$ & 0.000e+00 & (DD) & 6.603e-05 & (MD) & 259 & 10047\\
0.754 & 55930.9079 & $-129\pm26$ & $4\pm12$ & 0.000e+00 & (DD) & 7.283e-08 & (DD) & 275 & 10927\\
0.838 & 55931.1392 & $-42\pm14$ & $-27\pm18$ & 5.660e-08 & (DD) & 3.593e-02 & (ND) & 242 & 9306\\
0.843 & 55933.9247 & $-18.2\pm9.2$ & $11\pm13$ & 3.844e-11 & (DD) & 2.778e-02 & (ND) & 284 & 11603\\
0.919 & 55934.1345 & $72\pm17$ & $-23\pm19$ & 0.000e+00 & (DD) & 6.420e-02 & (ND) & 266 & 10596\\

    \hline
    \end{tabular}
    \tablefoot{$\langle B_z\rangle$ magnetic field measurements of each component during each observation. Also included is the FAP. DD, MD, and ND stand for definite-, marginal- and non-detection as defined by \cite{donati:1997}. A $^*$ indicates phases where the radial velocity windows employed to calculate the \cla{$\langle B_z \rangle$} and FAP of the two components overlapped, resulting in possible cross-talk of the FAP and $\langle B_z\rangle$ between the components.}
    
\end{table*}
\newpage
\section{Small-scale posterior distributions}
\begin{figure*}[h!]
    \centering
    \includegraphics[width=\textwidth]{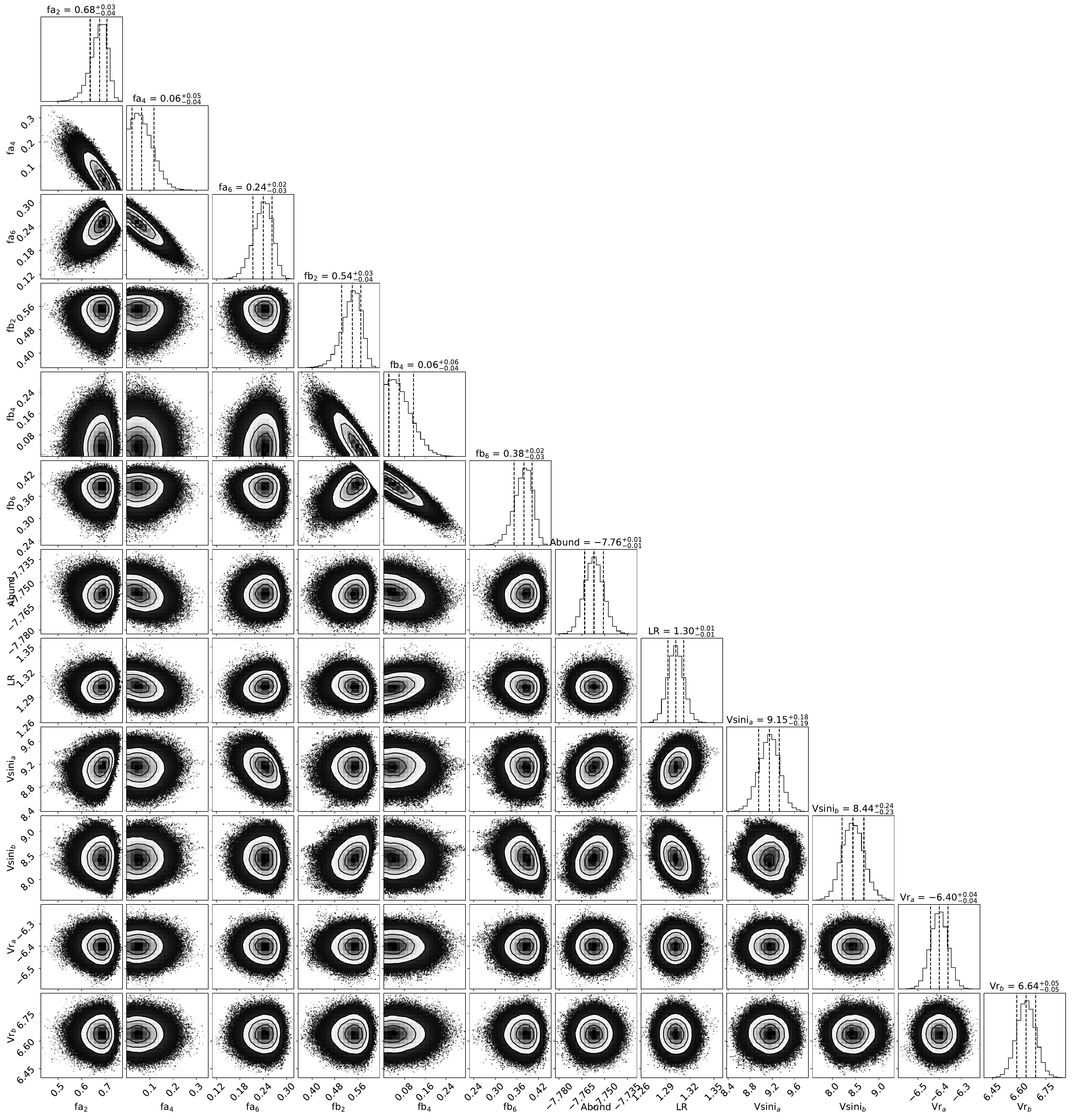}
    \caption{Corner plot from the small-scale magnetic field investigation containing the magnetic field filling factors as well as other free parameters for the two components.}
    \label{fig:small_scale_corner}
\end{figure*}
\end{appendix}
\end{document}